\documentclass[iop]{emulateapj}

\begin{document}

\title{Characterizing the Variability of Stars with Early-Release Kepler Data}

\author{David R. Ciardi\altaffilmark{1}, Kaspar von~Braun\altaffilmark{1},
Geoff Bryden\altaffilmark{2}
Julian~van~Eyken\altaffilmark{1}, Steve B. Howell\altaffilmark{3},  Stephen
R. Kane\altaffilmark{1}, Peter~Plavchan\altaffilmark{1},
Solange V. Ram\'{i}rez\altaffilmark{1}
John R. Stauffer\altaffilmark{4}}

\altaffiltext{1}{NASA Exoplanet Science Institute/Caltech Pasadena, CA 91125 USA}
\altaffiltext{2}{Jet Propulsion Lab/Caltech, Pasadena, CA 91109, USA}
\altaffiltext{3}{NOAO, 950 North Cherry Avenue Tucson, AZ 85719, USA }
\altaffiltext{4}{Spitzer Science Center/Caltech, Pasadena, CA 91125, USA}



\slugcomment{Accepted for publication in The Astronomical Journal}

\begin{abstract}

We present a variability analysis of the early-release first quarter of
data publicly released by the Kepler project.   Using the stellar
parameters from the Kepler Input Catalog, we have separated the sample into
129,000 dwarfs and 17,000 giants, and further sub-divided the luminosity
classes into  temperature bins corresponding approximately to the spectral
classes A, F, G, K, and M.  Utilizing the inherent sampling and time
baseline of the public dataset (30 minute sampling and 33.5 day baseline),
we have explored the variability of the stellar sample.  The overall
variability rate of the dwarfs is 25\% for the entire sample, but can reach
100\% for the brightest groups of stars in the sample. G-dwarfs are found
to be the most stable with a dispersion floor of $\sigma \sim 0.04$ mmag.
At the precision of Kepler, $>95$\% of the giant stars are variable with a
noise floor of $\sim 0.1$ mmag, 0.3 mmag, and 10 mmag for the G-giants,
K-giants, and M-giants, respectively. The photometric dispersion of the
giants is consistent with acoustic variations of the photosphere; the
photometrically-derived predicted radial velocity distribution for the
K-giants is in agreement with the measured radial velocity distribution.
We have also briefly explored the variability fraction as a function of
dataset baseline (1 - 33 days), at the native 30-minute sampling of the
public Kepler data.  To within the limitations of the data, we find that
the overall variability fractions increase as the dataset baseline is
increased from 1 day to 33 days, in particular for the most variable
stars.  The lower mass M-dwarf, K-dwarf, G-dwarf stars increase their
variability more significantly than the higher mass F-dwarf and A-dwarf
stars  as the time-baseline is increased, indicating that the variability
of the lower mass stars is mostly characterized by timescales of weeks
while the variability of the higher mass stars is mostly characterized by
timescales of days.  A study of the distribution of the variability as a
function of galactic latitude suggests sources closer to the galactic plane
are more variable.  This may be the result of sampling differing
populations (i.e., ages) as a function of latitude or may be the result of
higher background contamination that is inflating the variability fractions
at lower latitudes.  A comparison of the M dwarf statistics to the
variability of 29 known bright M dwarfs indicates that the M dwarfs are
primarily variable on timescales of weeks or longer  presumably dominated
by spots and binarity.  But on shorter timescales of hours which are
relevant for planetary transit detection, the stars are significantly less
variable, with $\sim 80$\% having 12-hour dispersions of 0.5 mmag or less.

\end{abstract}


\section{Introduction}\label{intro-sec}

Stars have been known for a long time to vary in brightness, and
photometric studies over the past centuries have revealed many classes of
stars exhibiting a variety of variability \citep{pickering1881}.  With
interest in stellar variability growing tremendously in the last decade as
ground-based and space-based surveys for exoplanets have gained momentum,
understanding the stellar photometric variability is even more crucial.

Sources of stellar variability include pulsations, binarity, rotation, and
activity \citep[e.g.,][]{em08}.  Having a large sample of uniformly
observed stars is vital in the categorization and characterization of the
variability which can inform us about the stars themselves, their
companions and companion rates, and their evolution.   The fractions of
stars that are found to be variable is dependent upon the sample of stars
studied, the precision of the survey, the range of magnitudes over which
the precision is matched, and the time duration of the survey \citep[e.g.,
][]{em08,howell08}.  For example, Hipparcos (with mmag precision and a
completeness limit near V=8 mag) found 10\% of the stars in the sample to
be variable \citep{eg97,em08}, but the variability fraction depended upon
both the stellar brightness and  the stellar type.  Similar population and
precision dependent results have been found by survey programs intended for
other purposes such as microlensing studies and transit surveys (e.g.,OGLE
\citep{ws98}, HATNet \citep{hartman04}, and  WASP0 \citep{kane05}) as well
as from general variability programs (e.g., BSVS \citep{everett02}, FSVS
\citep{heh06}, and ASAS \citep{pojmanski02}).

As the surveys have become more sensitive, the fraction of stars observed
to vary has been found to increase in a form which can be described by a
power-law distribution directly proportional to the quality of the
photometric precision \citep{howell08}.  This is a result of the current
``best'' survey precisions, time samplings, and survey durations probing
ever deeper into the variability of stars but generally not reaching the
astrophysical variability floor.

Spaced-based missions such as MOST \citep{matthews99}, CoRoT
\citep{auvergne09}, and Kepler \citep{borucki10} take advantage of the
controlled environment in space to achieve the best possible precision for
the telescope -- increasing the precision of the photometry and allowing us
to explore the limits of stellar variability.  The Kepler mission, with its
large 1m aperture and huge focal plane ($\sim 100^{\square^\circ}$), is
obtaining sub-millimagnitude precision (30 minute integration) and
micro-magnitude precision (6 hour timescale) for thousands of stars and has
the potential to expand our knowledge of the limits of stellar variability.

Kepler was launched in March 2009 and began science operations in May 2009.
Like CoRoT, Kepler does not study all stars within its field-of-view, but
rather Kepler monitors a specific set of $\sim 150,000$ target stars
\citep{batalha10}.  Early work on the variability of stars in the Kepler
dataset has been performed; these works have concentrated on the dwarf
stars, periodicity, and flares  \citep{basri10a, basri10b, walkowicz10}.

In June 2010, the Kepler project released to the public the first major
time series data product for the majority of the targets.   We present a
discussion of the dataset (\S~\ref{q1-sec})  and how it it is divided into
spectral and luminosity classes (\S~\ref{sampleseg-sec}).  We primarily
discuss the stellar variability of the sample on the time scale of the
dataset (33 days) and at the sampling rate of the data (30 minutes); we do
explore briefly the variability as a function of the time baseline from 1 -
33 days.   Discussions of the stellar photometric dispersions
(\S~\ref{dispersion-sec}) and the variability fractions
(\S~\ref{fraction-sec}) for the dataset as a whole (30-minute sampling,
33.5 day baseline) are presented.  The variability study is extended by
exploring the source of the variability in the giant stars, the
time-dependency of the variability fractions, and the variability fraction
as a function of galactic distribution (\S~\ref{glat-sec}).  Finally, we
explore in more detail the variability of the lower mass main sequence
stars (\S~\ref{mdwarf-sec}). Studies and characterization of stellar
variability not only provide insight into the nature of stars themselves,
but also help inform our statistical understanding of the detection of
transiting exoplanets in the presence of stellar ``noise''.

\section{Kepler Public Data}\label{keplerdata-sec}
\subsection{Quarter 1 and Characterization}\label{q1-sec}

The Kepler project publicly released light curve data for all targets
observed in the first two ``quarters'' of observing (Q0 and Q1) and for
targets listed by the Kepler project as ``dropped'' from observation in
quarters Q0, Q1, and Q3.  We have chosen to utilize only the  Q1 data for
this study, as these data represent the most complete and most uniform set
of Kepler data available to the public.  The Q1 data mark the beginning of
science operations and span approximately 33.5 days from the end of Q0 (13
May 2009) to first spacecraft roll (15 June 2009)\footnote{Data released to
the public June 2010.}.  We have also chosen to use only the 30 minute
cadence data (and not the 1 minute cadence data) to maintain the uniformity
and continuity of the sample.  The data are available through the Kepler
mission archive at MAST\footnote{http://archive.stsci.edu/kepler} and also
through the NASA Star and Exoplanet Database
(NStED).\footnote{http://nsted.ipac.caltech.edu}

In addition to providing access to the light curve data themselves, NStED
calculates a standard set of statistics for each light curve as a whole (33
day baseline at 30 minute sampling) including a median value, a median of
the uncertainties, a dispersion about the median value, and a reduced
chi-square assuming a constant (median) value.  The statistics are provided
as part of the header information in the NStED ASCII versions of the public
FITS files, and are also searchable and downloadable as part of the NStED
data query service. These statistics are calculated on the data corrected
by the Kepler project for ``instrumental effects'' (ap\_corr\_flux).  As
mentioned in the Kepler Data Release Notes \citep{vancleve10}, the Kepler
project is in the early development stages of the data processing pipeline,
which is primarily intended to find exoplanetary transits.  The  pipeline
may not perfectly preserve general stellar variability with amplitudes
comparable to or smaller than the instrumental systematics on long
timescales.

The Kepler project warns that trends in the data comparable to the length
of the time-series data ($\sim20-30$ days in the case of the Q1 data) may
not be fully preserved in the Kepler pipeline processing
\citep{vancleve10}. That is not to say that all long-term trends are
removed from the data by the Kepler processing, but the variability
statistics provided by NStED (and used in this study) are more sensitive to
variability shorter than a few weeks.   The primary effect of the Kepler
pipeline is over-correction for shorter datasets (like the Q0 data) and
fainter stars, but the pipeline is also capable of adding or enhancing
variability within the light curves \citep{vancleve10}.

Because we are interested in the overall variability statistics of the
sample and not in the variability or periodicity of any one individual
star, the sheer size of the sample ($\sim150,000$ stars) helps alleviate
the specific effects of any one star.  In addition, the variability
statistics presented in this work are in reasonable agreement with
statistics presented by \citet{jenkins10} and \citet{vancleve10}, and also
in reasonable agreement with the variability statistics of \citet{basri10a,
basri10b}, who use a ``range'' of variability to describe the statistics.
However, the results presented here should be viewed as a preliminary
exploration of the public data set and are subject to revision as the
Kepler project matures and improves the data products.

\subsection{Sample Segregation}\label{sampleseg-sec}

To help understand the variability statistics, we have utilized the Kepler
Input Catalog \citep[KIC; ][]{latham05,batalha10} to separate the stars
into broad spectral and luminosity classes. The KIC includes stellar
parameters (temperature and surface gravity) derived from photometric
observations ($u,g,r,i,z,DDO51,J,H,Ks$); a ``Kepler Magnitude''
corresponding to the bandpass of the instrument is derived from the
ground-based photometry \citep{koch10}. The primary purpose of the KIC was
to identify F, G, and K (and M) dwarfs and separate them from the
background giants in the field by utilizing photometry to determine
line-of-sight extinction, effective temperatures, and surface gravities
(see \citet{batalha10} for a description of the KIC algorithms and target
selection process).  These derived values are available as part of the KIC
information attached to each Kepler time series file.  Of the 152,919 light
curves available,  143,221 stars have KIC temperatures and surface
gravities which we have used to separate the sample into dwarfs and giants
by surface gravity and into spectral classes by temperature.

The KIC temperatures and surface gravities are based upon isochrone fitting
utilizing the ATLAS9 models \citep{batalha10}.  The KIC survey utilized the
DDO51 filter which is sensitive to the MgH+Mgb line strength which varies
as a function of surface gravity for G and K stars \citep{majewski00}.
\citet{basri10b} showed that the KIC did a reasonably good job of
separating giants from dwarfs, particularly for the G and K stars which
dominate the sample.

\begin{figure}

   \includegraphics[angle=0,scale=0.4,keepaspectratio=true]{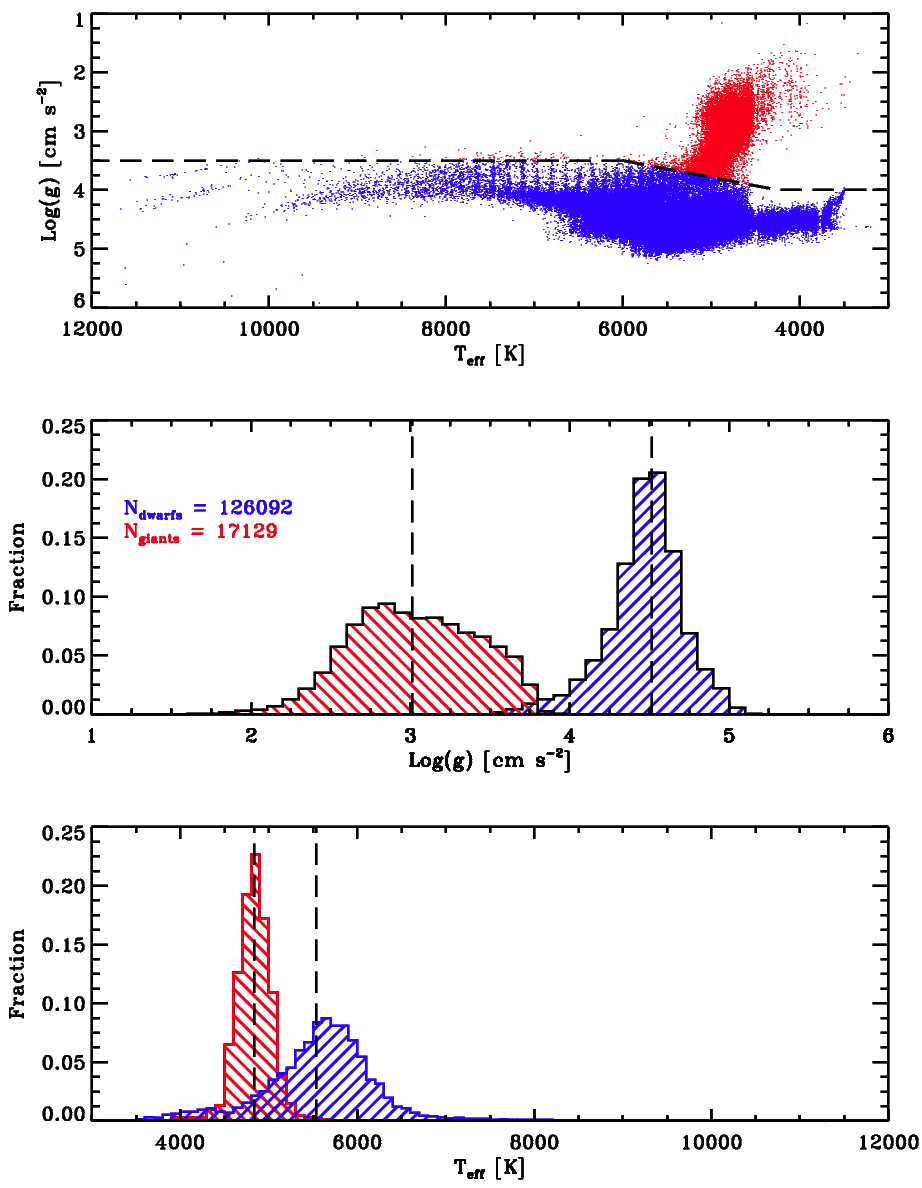}

    \figcaption{{\it Top}: KIC-based Surface Gravity -- Effective
    Temperature HR-diagram of the stars in the analysis sample.  The dashed
    black line marks the delineation to separate dwarfs (blue) and giants
    (red).  {\it Center}: Histograms of the surface gravity for the dwarfs
    (blue) and giants (red).  The vertical dashed lines mark the median
    surface gravity values. {\it Bottom}: Histograms of the effective
    temperatures for the dwarfs (blue) and giants (red).  The vertical
    dashed lines mark the median temperature
    values.\label{selectgiants-fig}}

\end{figure}

Separating the dwarfs and giants with a single value of surface gravity was
not found to be sufficient.  For example, a single surface gravity cut at
$\log(g)=4.0$ produces a bimodal distribution of the surface gravities for the
giant star distribution and a truncated tail for the dwarf distribution of
surface gravities; these artificial structures in the distributions
indicated that the giant sample was significantly contaminated by dwarf
stars at the 20\% level.  In an effort to transition more naturally between
giants and dwarfs, we have employed a three-section (empirical) surface
gravity cut determined from the surface gravity-effective temperature HR
diagram (see Figure \ref{selectgiants-fig}). For three separate temperature
ranges, a star was considered to be a dwarf if the surface gravity was
greater than the value specified in the following algorithm:
\[
{\rm log(g)} \geq \left\{ \begin{array}{lll}
						   3.5 & &{\rm if\ } T_{eff} \geq 6000\\
                           4.0 & &{\rm if\ } T_{eff} \leq 4250\\
                           5.2-(2.8\times10^{-4}T_{eff}) &
                           & {\rm if\ } 4250 < T_{eff} < 6000\\
                           \end{array}
                   \right.
\]
The delineation between dwarfs and giants is shown in
Fig.~\ref{selectgiants-fig} by the dashed line with the dwarfs and giants
highlighted in blue and red, respectively.

\begin{figure}

    \includegraphics[angle=0,scale=0.4,keepaspectratio=true]{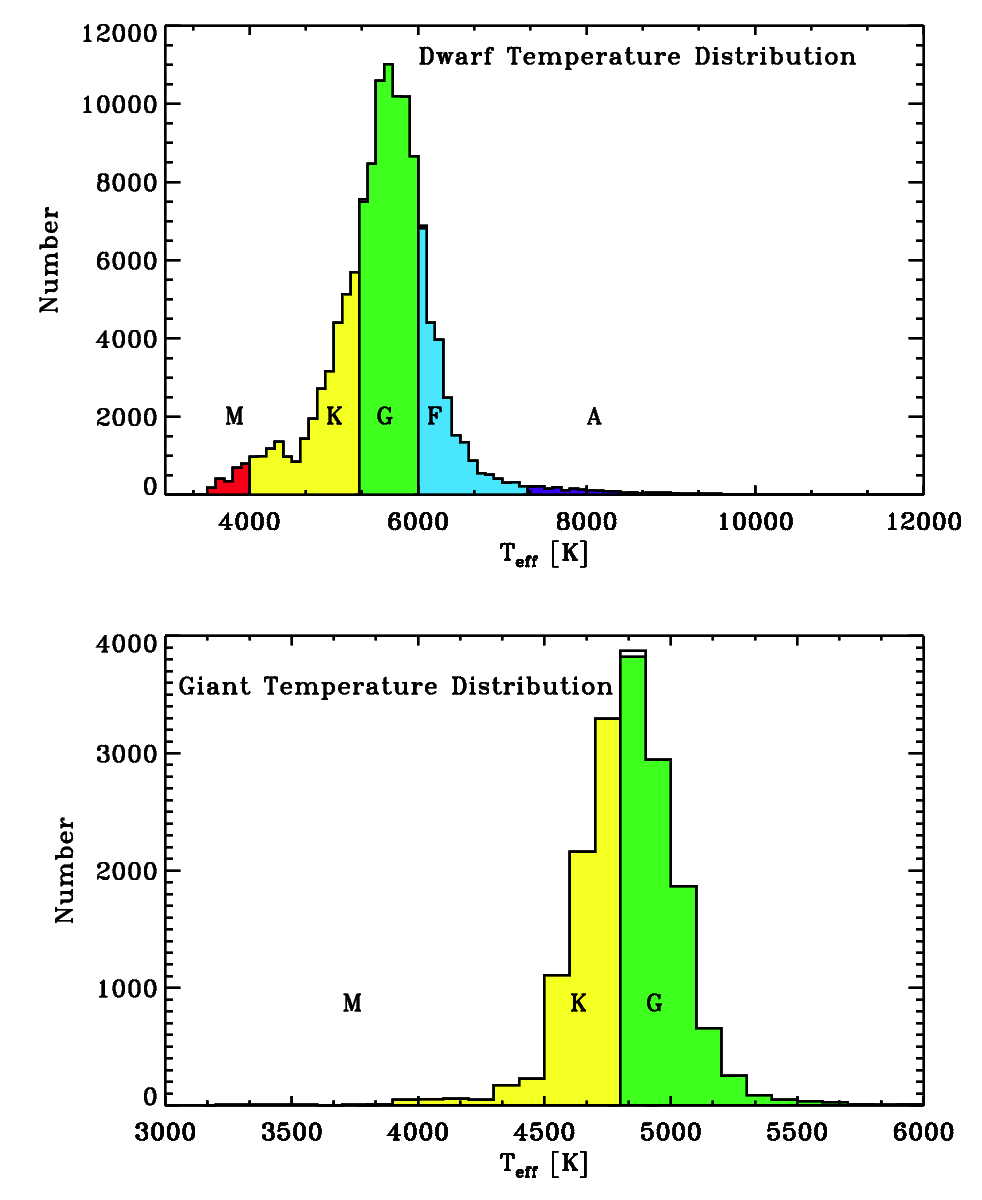}

    \figcaption{Temperature distributions (binsize = 100 K) of the stars
    selected as dwarfs ({\it top}) and giants ({\it bottom}).   The
    color-coding illustrates the separation of the dwarfs and giants into
    temperature groups (i.e., spectral types).\label{selecttemp-fig}}

\end{figure}

\begin{figure}

    \includegraphics[angle=0,scale=0.8,keepaspectratio=true]{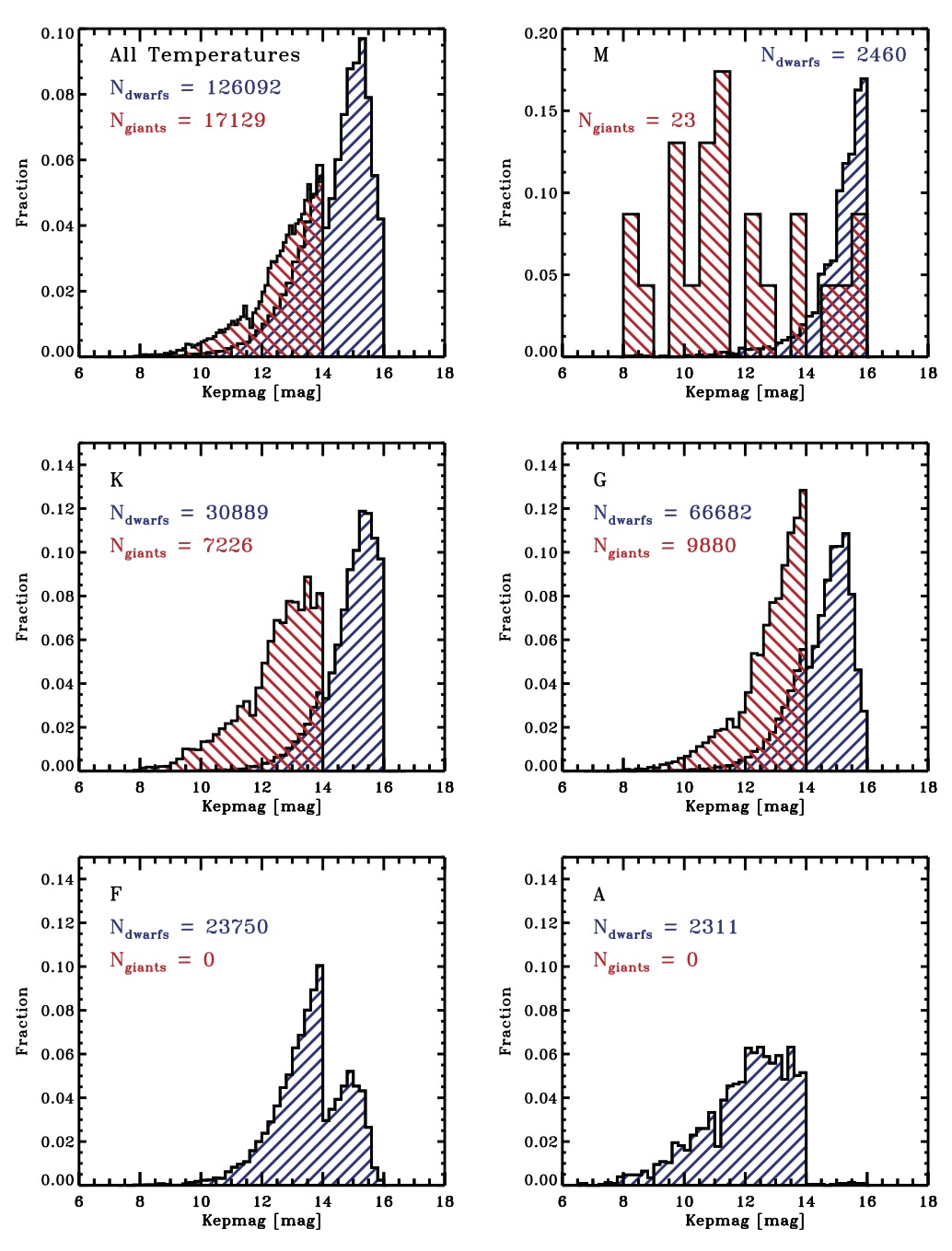}

    \figcaption{Kepler magnitude distributions of the stars in the sample.
    Dwarfs and giants are represented by the blue and red hashed
    histograms, respectively. The panels represent the different
    temperature groups as labeled in the figures.\label{magdist-fig}}

\end{figure}

The total number of stars separated into dwarfs and giants are 126,092 and
17,129, respectively.  There is clear separation in the distributions of
surface gravity for the two groups of stars (see middle panel
Fig.~\ref{selectgiants-fig}).  The median surface gravities for the dwarfs
and giants are, respectively, $\log(g)=4.5$ and $\log(g)=3.0$ with a small
overlap in surface gravity near $\log(g)=3.7$.  The overlap is likely
dominated by sub-giants but represents a small contamination rate for
both the dwarf and giant samples.  The temperature distributions of the
stars are shown in the lower panel of Fig.~\ref{selectgiants-fig}; the
median dwarf and giant temperatures are 5500 K and 4800 K, respectively.
The dwarfs and giants have further been separated into temperature bins
corresponding roughly to the spectral types A, F, G, K, and M
\citep{johnson66,dl00}; the temperature binning for each spectral class is
listed in Table~\ref{temp-tab} and illustrated in
Figure~\ref{selecttemp-fig}. The numbers of A and M stars are relatively
small in comparison to the F, G, and K stars, but are maintained in the
study for completeness.  The temperature distributions clearly show that
the G and K stars (and F dwarfs) dominate the sample.  The magnitude
distributions of the stars, separated by temperature and by dwarfs and
giants, are shown in Figure~\ref{magdist-fig}.

\begin{figure}

    \includegraphics[angle=0,scale=0.45,keepaspectratio=true]{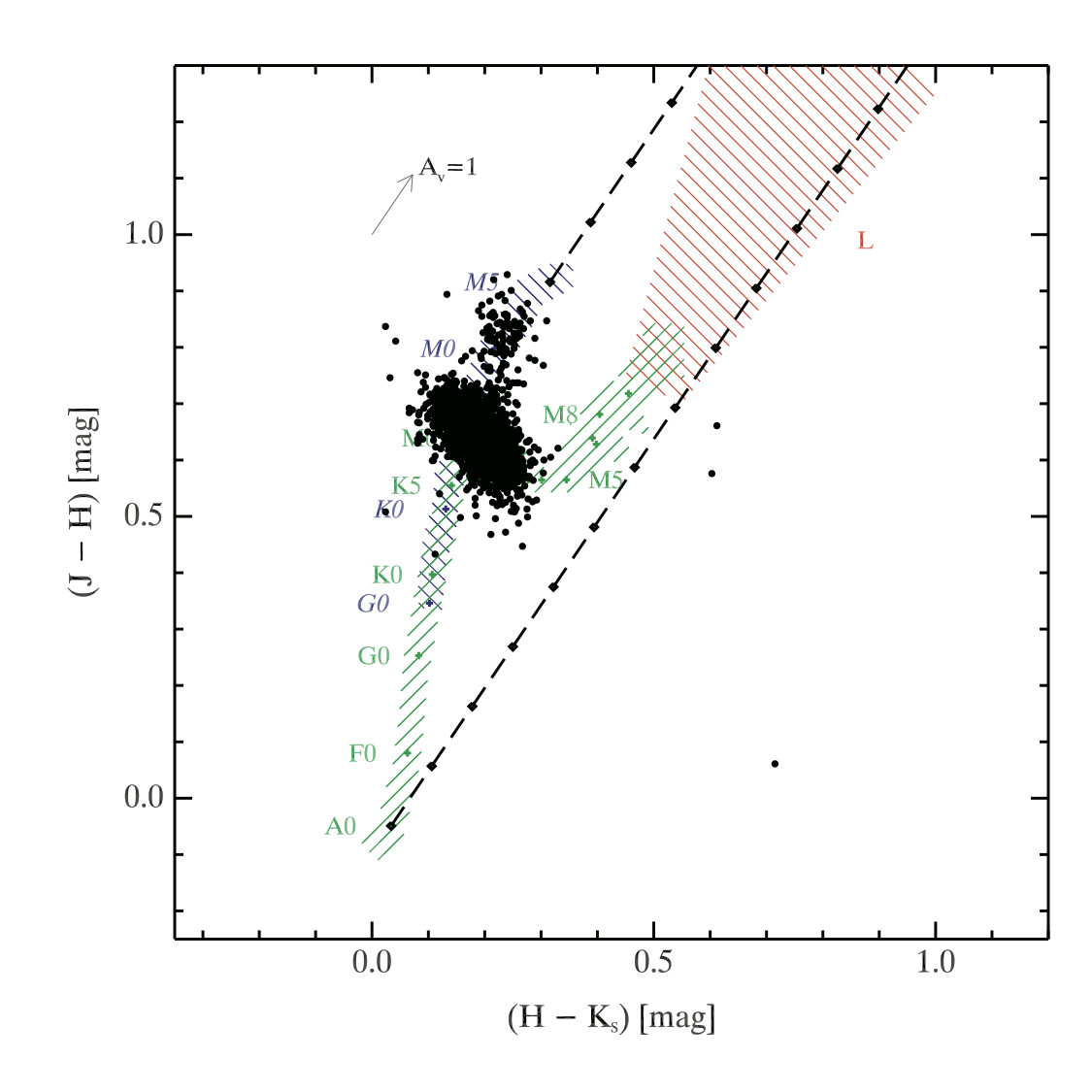}
    \includegraphics[angle=0,scale=0.45,keepaspectratio=true]{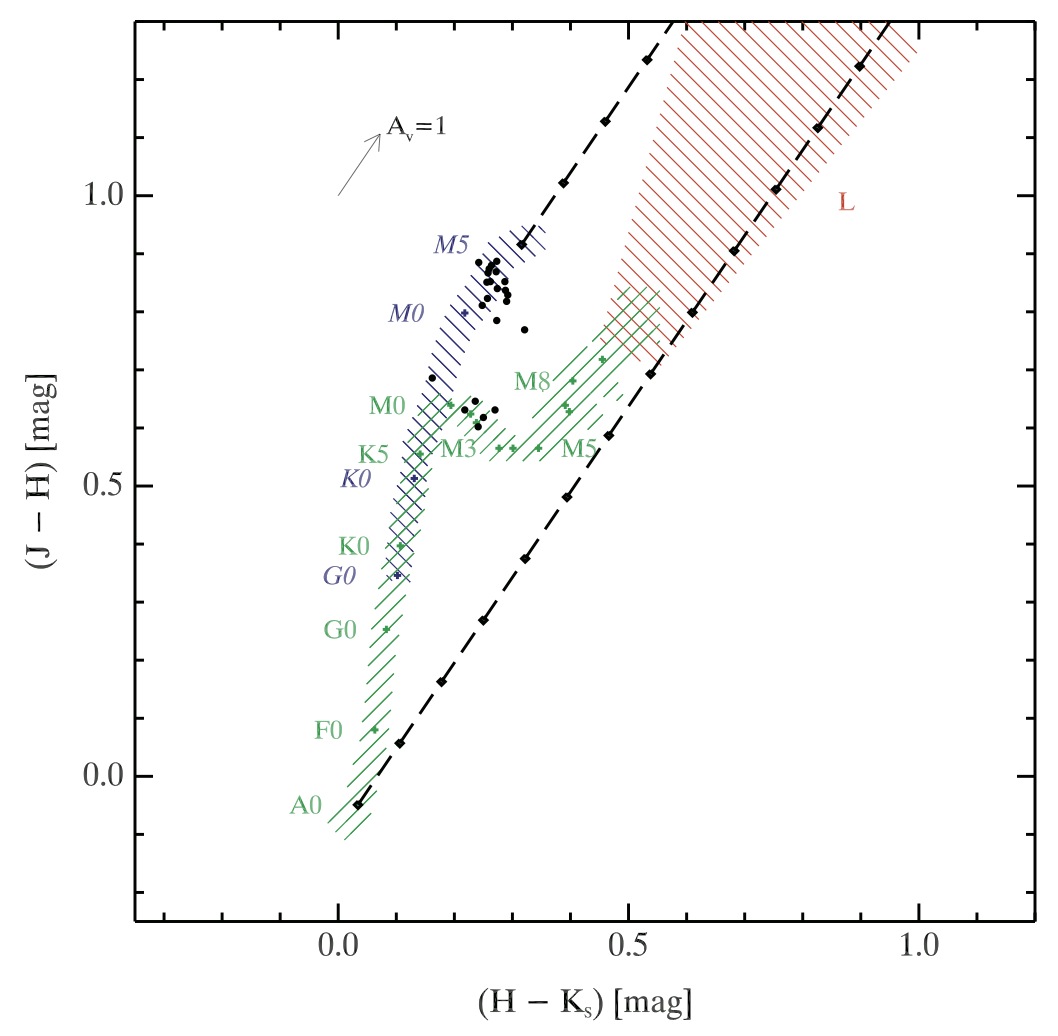}

    \includegraphics[angle=0,scale=0.45,keepaspectratio=true]{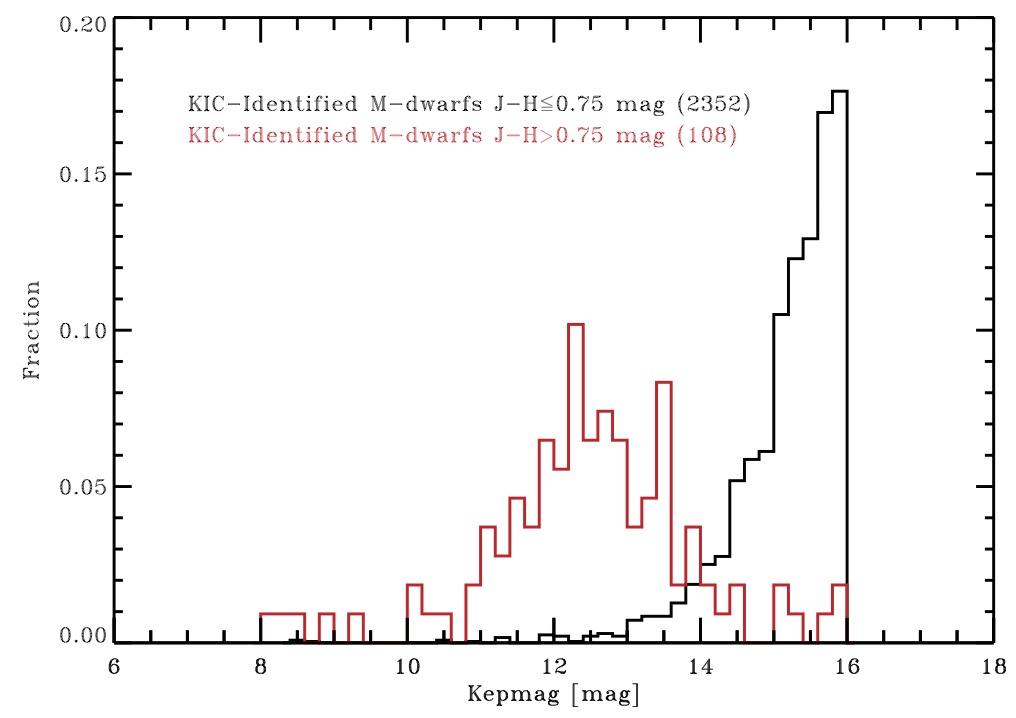}
    \includegraphics[angle=0,scale=0.45,keepaspectratio=true]{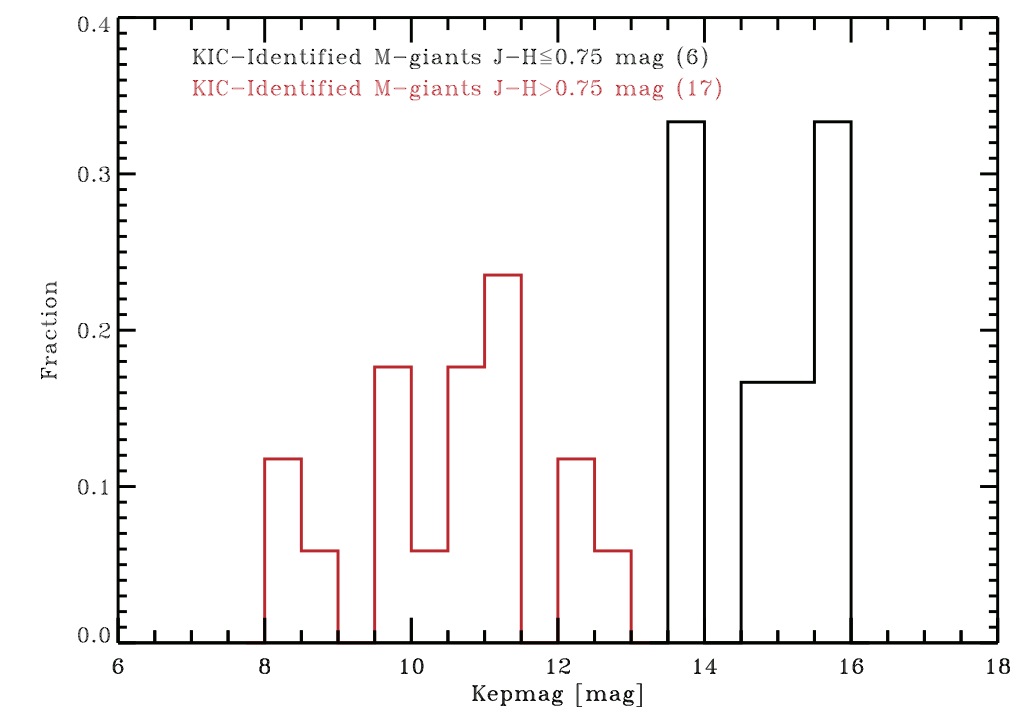}

    \figcaption{{\it Top}: 2MASS color-color diagram for the stars
    identified as M-dwarfs (left) and M-giants (right) based upon the KIC
    surface gravities and effective temperatures.  The green-hashed area
    marks the main sequence; the blue-hashed area marks the giant branch,
    and the red-hashed area marks the L-dwarf locus. The diagonal lines
    mark the reddening zone for typical galactic interstellar extinction
    (R=3.1). {\it Bottom}: Magnitude distributions for stars identified as
    dwarfs (left) and as giants (right) by their surface gravity.  The
    black histograms are for stars with dwarf-like J-H colors; the red
    histograms are for stars with giant-like J-H colors.  These plots show
    that the M-stars brighter $\sim$Kepmag$< 13.5$ mag are predominately
    giants, regardless of their KIC classification. \label{mjhk-fig}}

\end{figure}

We have specifically explored the contamination rate of the M dwarfs with
giant stars, by placing the M dwarfs on a 2MASS JHKs color-color diagram,
where the dwarf and giant colors are sufficiently different to enable
separation (see Figure \ref{mjhk-fig}).  Note that all of the M-dwarfs, as
identified from the KIC, have surface gravities of $\log(g)>4$; yet, it is
clear from the color-color diagram that a fraction of those identified as
dwarfs are indeed giants. Using $J-H = 0.75$ mag as the boundary between
dwarfs and giants, we find that only $\approx4\%\ (108/2460)$ of the entire
sample of stars identified as M-dwarfs in the KIC actually have infrared
colors of a giant star. However, these contaminating stars are
overwhelmingly brighter than the general M-dwarf sample with 80\% of the
giant-color ``dwarfs'' having a Kepmag brighter than 13.5 mag (see
Fig.~\ref{mjhk-fig}).  Thus, at the bright-end of the M-dwarf sample
(Kepmag$\lesssim 13.5$ mag),  the giant contamination rate is $\gtrsim50\%$
(87/170).   The inverse contamination is also evident.  The entire M-giant
sample is much smaller with only 23 stars in total, but, of these, 6
($\sim$25\%) have JHK colors of dwarfs.  The contaminating M dwarfs are
systematically fainter than the true M-giants.  For the sake of uniformity
and continuity, we have not moved the contaminating sources into
corresponding ``correct'' category; we do, however, exclude them when
calculating the variability fractions.

\section{Variability}\label{variability-sec}

For the spectral and luminosity classes defined above, we have assessed
the distributions of the dispersion and variability to understand the broad
stellar variability characteristics across the stellar spectrum. The
analysis presented here utilized the statistics provided by NStED where the
data were assessed using the native 30-minute sampling and the full 33-day
time baseline  of the dataset. The time series data are characterized by
the dispersion about the median ($\sigma_m$) and by the reduced chi-square
assuming a constant median value for the light curve ($\chi^2_\nu$).  The
first part of the study discusses the measured dispersions, and the second
part of the study discusses the variability fractions of the stars within
each group of stars.  We also briefly explore the variability fraction of
the stars as a function of the time baseline of the dataset (1 - 33 days).

\subsection{Photometric Dispersion}\label{dispersion-sec}

\begin{figure}

    \includegraphics[angle=0,scale=0.6,keepaspectratio=true]{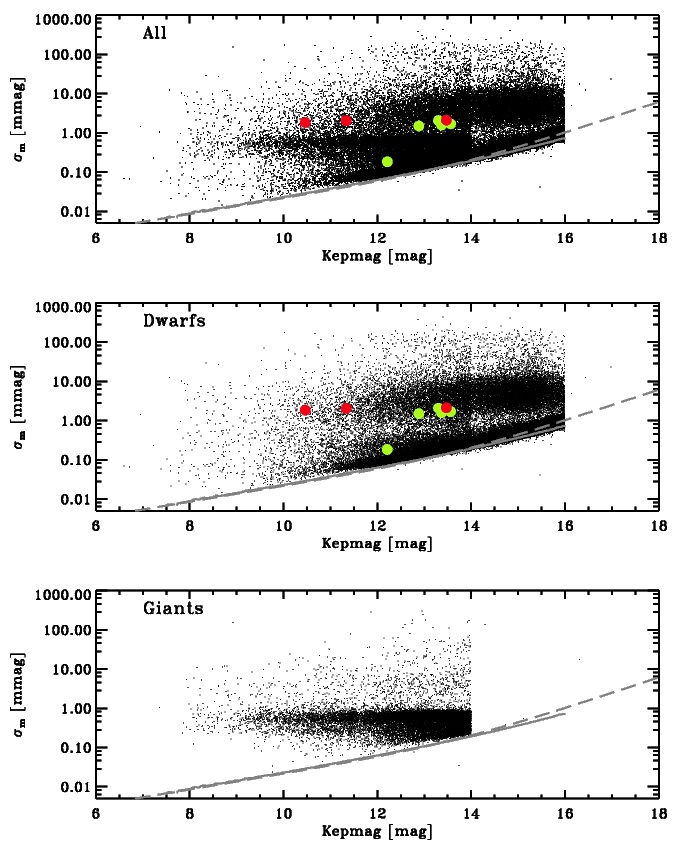}

    \figcaption{Photometric dispersion (30 minute sampling; 33 day
    timescale) of each star is plotted as a function of magnitude for all
    stars (top), for just the  dwarfs (center), and for just the giants
    (bottom). In the top and center plots, the locations of the 7 known
    planets in the sample  are shown (red: BOKS-1, Hat-P7, and TrES-2;
    green: Kepler-4,5,6,7,8).  The grey line represents the median
    uncertainty as reported in the Kepler data product. The dashed grey
    curve is the uncertainty upper limit curve from \citet{jenkins10}.
    \label{sigmag-fig}}

\end{figure}

\begin{figure}

    \includegraphics[angle=0,scale=0.8,keepaspectratio=true]{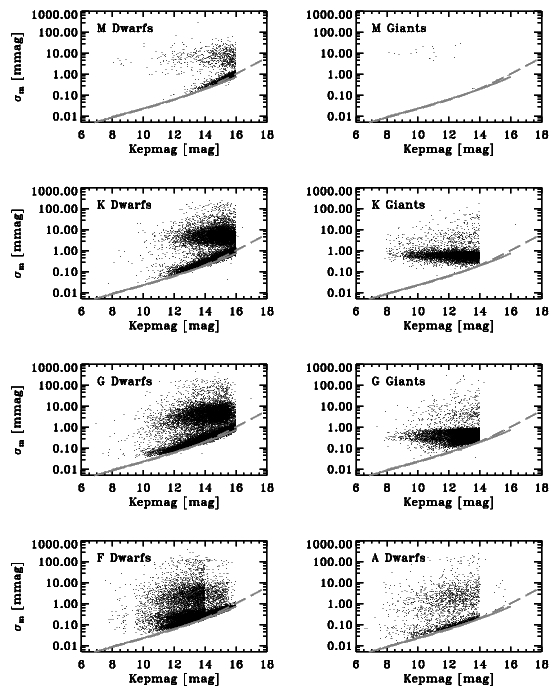}

    \figcaption{Photometric dispersion (30 minute sampling; 33 day
    timescale) of each star is plotted as a function of magnitude separated
    out by temperature and surface gravity as labeled in each panel  The
    solid grey line represents the median uncertainty value as reported in
    the Kepler data product. The dashed grey curve is the uncertainty
    upper limit curve from \citet{jenkins10}. \label{sigmagsep-fig}}

\end{figure}

Figure~\ref{sigmag-fig} shows the 30-minute, 33-day photometric dispersion
as a function of Kepler magnitude for all the stars and separated out by
dwarfs and giants, and Figure~\ref{sigmagsep-fig} displays the dispersions
to the same scale, but separated by temperature as well.    The grey dashed
lines in Figs.~\ref{sigmag-fig} and \ref{sigmagsep-fig} correspond to the
upper boundary on the uncertainties determined empirically for a constant
background component \citep[see $\sigma_{upper}$ in][]{jenkins10}. The grey
solid line represents the median uncertainty value as a function of Kepler
magnitude determined from the uncertainties provided with the data product
as part of the light curves.  To give some quantitative context to the
numbers of stars within Figs.~\ref{sigmag-fig} and \ref{sigmagsep-fig},
Table~\ref{disp-tab} tabulates the numbers of stars within four different
ranges of dispersion.  As instrumental precision plays a key role in the
dispersion of the stars (particularly at the faint end), the  stars are
grouped not only by stellar class but also by magnitude range.

\begin{figure}

    \includegraphics[angle=0,scale=0.4,keepaspectratio=true]{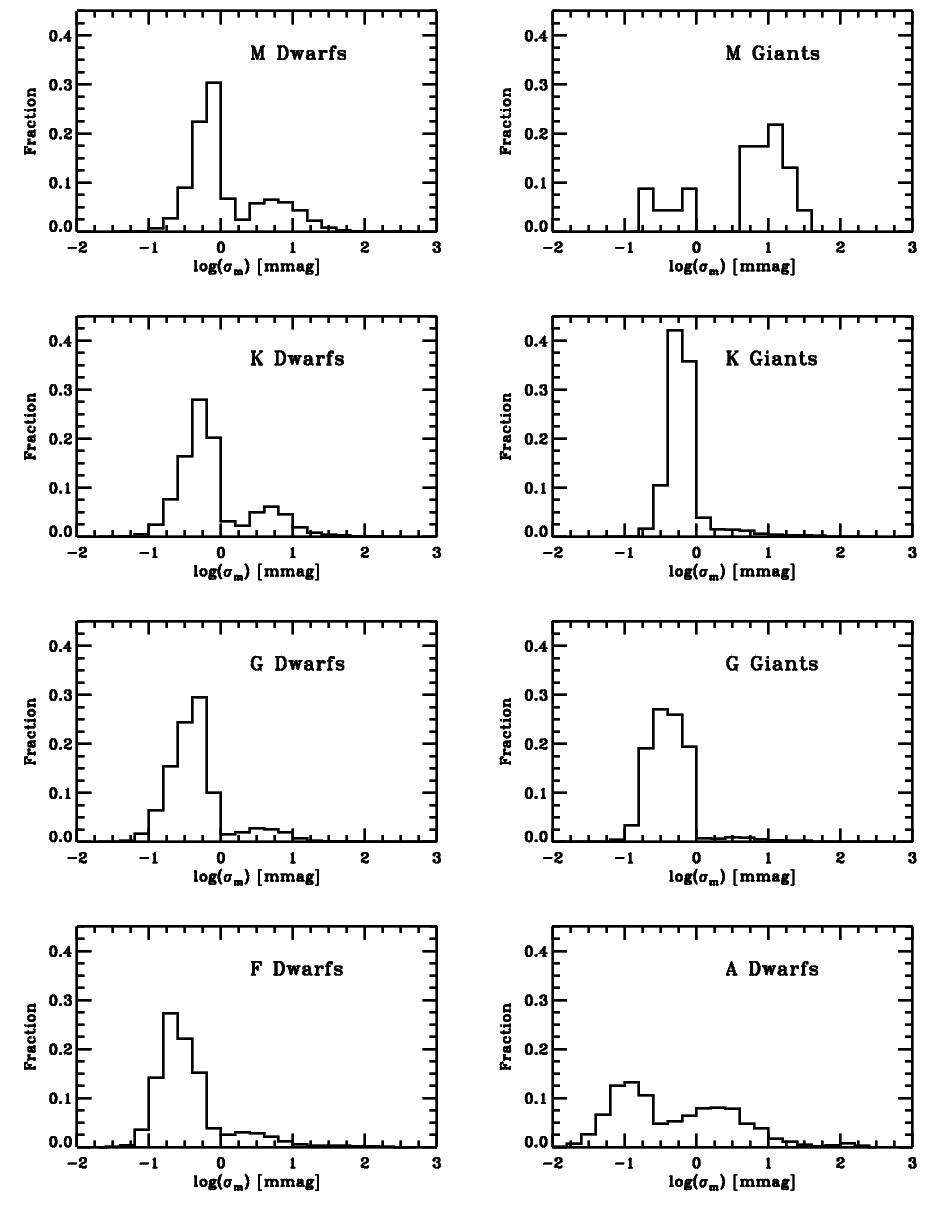}

    \figcaption{Distributions of the (logarithmic) photometric dispersion
    (30 minute sampling; 33 day timescale) separated out by effective
    temperature and luminosity class as labeled in each
    panel.\label{sigdist-fig}}

\end{figure}

There are a few specific aspects to the dispersion diagrams that are worth
noting.  At fainter magnitudes (Kepmag $\gtrsim 14$ mag), the model
uncertainties (grey solid and dashed lines in Figures \ref{sigmag-fig},
\ref{sigmagsep-fig}) track the stellar dispersion distribution fairly well
\citep[see also][]{jenkins10}. At brighter magnitudes (Kepmag $\lesssim 14$
mag), the model uncertainties track the lower bound of the measured
dispersions suggesting that Kepler is nearing the noise floor of the stars.
This effect is most clearly seen in the giant stars where the Kepler data
have sufficient precision to detect the floor of the variability for the
giant stars. The G- and K-giants occupy a very narrow range of photometric
dispersion between $0.1-1.0$ mmag - completely independent of the
magnitude.  This narrow range of dispersion is most clearly apparent in the
dispersion distribution histograms (Figure~\ref{sigdist-fig}).

\begin{figure}

    \includegraphics[angle=0,scale=0.5,keepaspectratio=true]{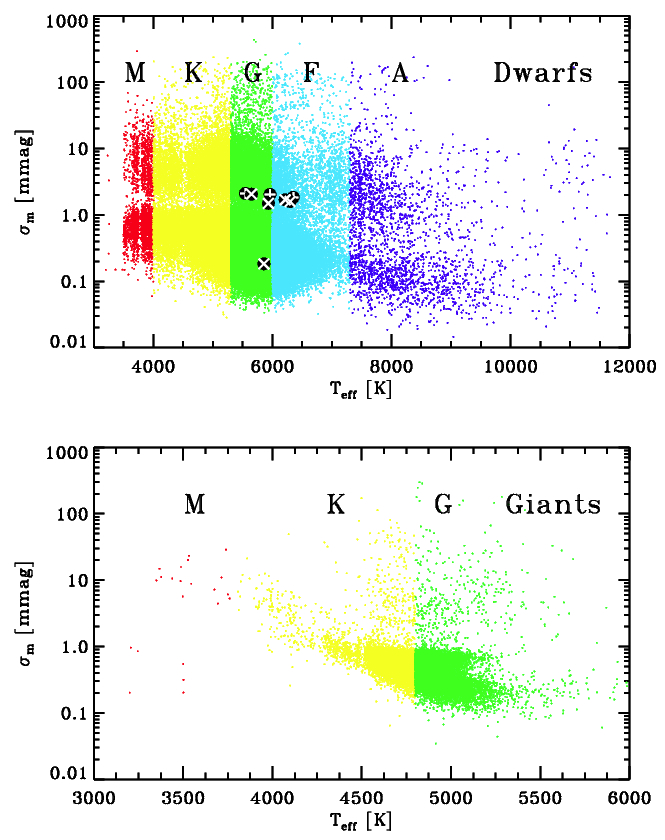}

    \figcaption{Photometric dispersion (30 minute sampling; 33 day
    timescale) of each star is plotted as a function of effective
    temperature separated out by temperature (colors and labels) and
    surface gravity (top and bottom panels).  The black points in the top
    panel mark the locations of the seven known planets in the sample ($+$:
    BOKS-1, HAT-P7, and TrES-2; $\times$:
    Kepler-4,5,6,7,8).\label{sigtemp-fig}}

\end{figure}

The ubiquity of variability in giants has been noted previously
\citep{gilliland08} for a set of galactic bulge stars observed by HST over
a time span of 7 days.  \citet{gilliland08} found the typical amplitudes
of variability was $\sim0.5$ mmag for the G giants and increased to $\sim
3.5$ mmag for the late-K to early-M giants.  We see a very similar trend in
the dispersion which is most clearly demonstrated in
Figure~\ref{sigtemp-fig} where we have plotted the photometric dispersion
as a function of effective temperature. While there is a scattering of
stars with large dispersions (and the number of M giants is very small),
the giant stars occupy a very narrow region of variability that is
correlated with temperature.  As expected from stellar evolution, the
larger, cooler giants are more variable \citep[e.g., ][]{kb95} and the
variability spans two orders of magnitude ($0.1-10$ mmag).

The dwarf stars are more complicated to interpret because their intrinsic
dispersion is on the order of (or less than?)~the photometric precision.
Taken as a whole, they are more quiescent than the giant stars, as expected
and demonstrated previously
\citep{gilliland08,jenkins10,vancleve10,basri10b}.  But there is a sample
of stars at all magnitudes (Fig.~\ref{sigmag-fig}) and all temperatures
(Figs.~\ref{sigmagsep-fig}, \ref{sigdist-fig}, \ref{sigtemp-fig}) where the
average dispersion is $\sim 5$ mmag.  Histograms of the dispersion
(Fig.~\ref{sigdist-fig}) and plotting the dispersion as a function of
temperature (Fig.~\ref{sigtemp-fig}) highlight the bi-modal dispersion, but
show that only a relatively small percentage of stars are in the higher
dispersion region.

Visual inspection of a sample of 50 light curves (10 light curves per
temperature bin) in the high dispersion region indicates that these light
curves are often periodic. Utilizing the NStED online periodogram
service\footnote{http://nsted.ipac.caltech.edu/applications/ETSS/kepler\_index.html}
$\sim90\%$ of the inspected light curves displayed one or more significant
periods (the origin and distribution of the periods were not explored in
this work).  A similar visual inspection of 50 stars in the lower
dispersion region (but flagged as variable with $\chi^2_\nu > 2$) revealed
that the variability was dominated by more stochastic ``white noise''
rather than periodic variability, and only $\sim 25\%$ of the stars
displayed significant periodicity. It is also possible that these higher
dispersions are an artifact of the  data processing; however, this bi-modal
dispersion distribution  (Fig.~\ref{sigdist-fig}) for the dwarfs is also
reported (although weaker) in \citet{basri10b} where they report a
``variability excess'' for those stars that are periodic versus those stars
that are not periodic (\citet{basri10b} independently processed the data
utilizing an empirical polynomial fitting process).  The details of the
variability (e.g., periodic or stochastic, amplitude, structure) have not
been fully explored in the work presented here, but it should be noted that
variability does not necessarily mean periodic behavior \citep{howell08} as
all the stars in the visual inspection were flagged as variable, but not
all stars were (obviously) periodic.

The Kepler light curves are precise enough that even small variations in
the light curve can lead to high dispersion, where in typical ground-based
transit survey data, the dispersion would remain relatively unchanged;
transiting Jupiter-sized planetary companions can significantly affect the
measured dispersion. To help put this into perspective, we have
over-plotted the positions of the known Kepler-field planets \citep[BOKS-1,
Hat-P7, TrES-2, Kepler-4,5,6,7,8;][] {howell10, pal08, odonovan06,
borucki10} on the dispersion diagrams
(Figs.~\ref{sigmag-fig},\ref{sigtemp-fig}).  The dispersions of the light
curves in these systems are $\sim 2$ mmag, except for Kepler-4 where the
light curve dispersion is $\sim 0.2$ mmag. These light curves are nearly
flat, to within the noise, except for the deep exoplanetary transits.  If
the transits are removed and the light curve statistics are recalculated,
the dispersions decrease by almost an order magnitude for all the light
curves except Kepler-4.  All the planets (except Kepler-4) are
Jupiter-sized with transit depths of $\sim 1\%$), and it is the transits
which dominate the statistics of the light curves. Kepler-4 is a much
smaller (Neptune-sized) planet with a transit depth of only $\sim 0.1\%$
which is comparable to the overall dispersion of the light curve.

\subsection{Variability Fractions}\label{fraction-sec}

\begin{figure}

   \includegraphics[angle=0,scale=0.5,keepaspectratio=true]{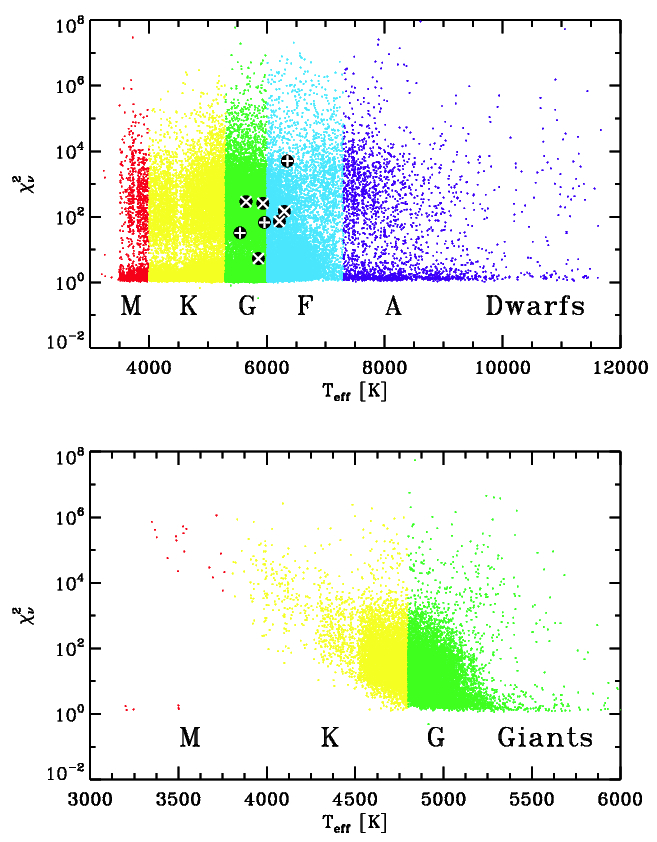}

    \figcaption{Photometric reduced chi-square (30 minute sampling; 33 day
    timescale) of each star is plotted as a function of effective
    temperature separated out by temperature (colors and labels) and
    surface gravity (top and bottom panels).  The black points in the top
    panel mark the locations of the seven known planets in the sample ($+$:
    BOKS-1, HAT-P7, and TrES-2; $\times$:
    Kepler-4,5,6,7,8).\label{chitemp-fig}}

\end{figure}

The photometric dispersions (33-day baseline, 30-minute sampling) alone are
not sufficient to assess the fraction of stars that are variable as the
dispersion is dependent on the apparent magnitude of the targets, and, in
particular, the dispersion for the dwarfs is at (or near) the precision
limits of the instrument (for the 30-minute cadence). A more natural
statistic is the reduced chi-square ($\chi^2_\nu$) which takes into account
the uncertainties (as reported in the public data light curve files). This
analysis makes use of the  provided point-to-point uncertainties in the
light curves.  For all light curves, the reduced chi-squares are calculated
for the 33-day baseline (30-minute sampling) with respect to a constant
median value and are plotted as a function of temperature
(Fig.~\ref{chitemp-fig}). For variability assessment purposes, a star is
considered just-barely variable if $\chi^2_\nu > 2$, significantly
variable if $\chi^2_\nu > 10$, and very variable if $\chi^2_\nu > 100$.  A
$\chi^2_\nu \approx 2$ corresponds to an excess dispersion of approximately
1.5 times that of the measurement uncertainties; a $\chi^2_\nu \approx 10$
corresponds to an excess dispersion of approximately 3 times that of the
measurement uncertainties, and a $\chi^2_\nu \approx 100$ corresponds to an
excess dispersion of approximately 10 times that of the measurement
uncertainties.

\begin{figure}

    \includegraphics[angle=0,scale=0.7,keepaspectratio=true]{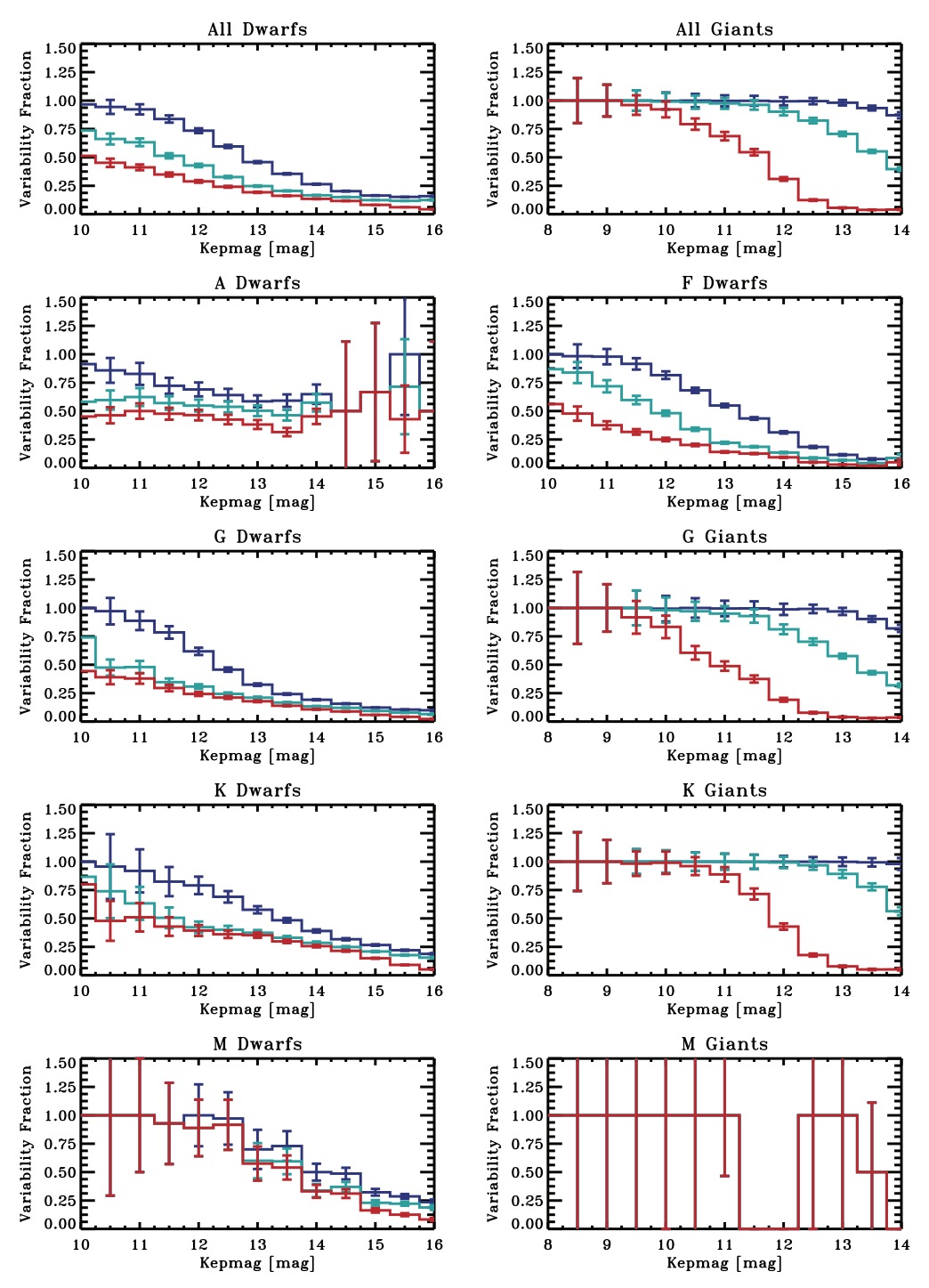}

    \figcaption{Variability fractions of stars as a function of the
    brightness (Kepler magnitude). The contaminating giants in the M-dwarf
    sample and the contaminating dwarfs in the M-giant sample have been
    removed from the statistics. The blue curves represent the fractions of
    stars with $\chi^2_\nu > 2$; the green curves represent the fractions
    of stars with $\chi^2_\nu > 10$;  the red curves represent the
    fractions of stars with $\chi^2_\nu > 100$. \label{varfracmag-fig}}

\end{figure}

The measured fractions of stars that are variable are dependent upon the
brightnesses of the stars as the instrumental precision decreases as the
stars become fainter.  Figure~\ref{varfracmag-fig} plots the variability
fractions as a function of the Kepler magnitude for each of the stellar
subgroups.  The magnitude bins are 0.5 magnitudes in width, and the
fractions were calculated for the three reduced chi-square categories
listed above.  Uncertainties on the fractions were calculated using
standard error propagation \citep{everett02}.  Some of the magnitude bins
have very few stars particularly at the bright end, and this is reflected
in the relatively large error bars.

At the very bright end of the Kepler sample (Kepmag $\lesssim 11$ mag), the
variability fractions for the stars with $\chi^2_\nu > 2$ are all near
unity indicating that the Kepler precision at 30-minute sampling is
approaching the 33-day noise floor for the stars.  Also for the brightest
stars, the fractions of stars that are significantly variable ($\chi^2_\nu
> 10$) is $50-100$\% depending on the sub-group of stars.  The fractions
decrease as the stellar magnitude increases; this is, of course, a  direct
result of the decrease in the instrumental precision as the stars become
fainter.   Not surprisingly, at all brightnesses, the fractions of stars
that are significantly variable ($\chi^2_\nu > 10 - 100$) are less than the
fraction of stars that are just-barely variable ($\chi^2_\nu > 2$).   For
the dwarfs, as the stars grow fainter (Kepmag $\gtrsim 14$ mag), the
variability fractions are typically dominated by the extremely variable
stars ($\chi^2_\nu > 10 - 100$), again a result of the lower precision on
the fainter stars.

A summary of the variability fractions is given in Table~\ref{var-tab},
where the fractions for each group of stars have been calculated for the
entire sample (Kepmag $< 16$ mag) and for the brighter end of the sample
(Kepmag $< 14$ mag).  The table includes variability fractions for the
whole light curve baseline (33-days) as well as for 1-day and 10-day time
baselines which are discussed in more detail in \ref{timevar-sec}.

\subsubsection{The Dwarf Stars}\label{dwarf-sec}

The G-dwarfs are the least variable group of dwarf stars with $>80\%$ of
the stars being stable (Kepmag $< 16$ mag; $\chi^2_\nu < 2$); even with the
magnitude restricted to the brightest stars (Kepmag $< 14$ mag), the
variability fraction of the G-dwarfs is $\sim 30$\%.  The floor for the
G-dwarf dispersion appears near 0.04 mmag (see Fig~\ref{sigmagsep-fig}).
The K-dwarf and F-dwarf stars have comparable variability fractions with
$\approx 50$\% of the stars identified as a variable if the magnitude is
restricted to Kepmag $< 14$ mag.  The F-dwarfs have a higher variability
fraction if F-dwarfs of all magnitudes are considered, but this is a result
of the different magnitude distributions of the K and F dwarfs in the
sample (see Fig.~\ref{magdist-fig}).  The relative number of F and K stars
brighter than and fainter than Kepmag $\approx 14$ mag differ, with the
K-dwarfs having significantly more fainter stars than brighter stars.   The
differing magnitude distributions for the F- and K-dwarfs is a result of
the target selection criteria optimized for searching for transiting
planets around as many appropriate stars as possible. \citep{batalha10}.

The M-dwarfs, not surprisingly, are less stable than the G, K, and F
dwarfs. If the magnitude is restricted to Kepmag $< 14$ mag, nearly 70\% of
the stars are variable (this is after the removal of stars which appear to
have giant star colors of $J-H > 0.75$ mag). The variability fraction drops
to 36\% if all the stars in the sample are considered. The A-dwarfs have a
similar variability fraction with $\approx  70$\% of the A-dwarfs being
variable.  Similar results are seen in the Hipparcos variability
statistics, where the A-dwarfs display the highest variability fractions
\citep[see Figure 2 of][]{em08}.  The large fraction of variable A-stars
is likely the result of the A-star group (as identified from the KIC)
including stars in the instability regime, such $\gamma$ Dor, $\delta$ Scu,
slowly pulsating B-stars (SPBs), RR Lyr, and  $\beta$ Cep stars.  Hipparcos
found the variability fractions of these sub-groups to range from $10 -
100$\%.

\subsubsection{The Giant Stars}\label{giant-sec}

At the precision of Kepler, nearly all of the giants are variable, with
94\%, 99\%, and 100\% variability fractions (33-day baseline, 30-minute
sampling) for the G, K, M giants, respectively. The 6 M-stars with dwarf
J-H colors have been removed from the statistics.  The G-giants variability
fraction is slightly reduced by the faint end of the brightness
distribution, where the stability floor approaches the instrument limit for
Kepmag $\approx 14$ mag, but only because the stability floor of the G
giants is 0.1 mmag versus 0.3 mmag for the K-giants.   For the M-giants,
the dispersion and stability floor is substantially higher at levels of
$\sim 10$ mmag.

The variability fraction of the giants found in the Kepler data is
consistent with the work of \citet{gilliland08} and \citet{em08}, where the
majority of the giants were found to be variable and a strong correlation
of variability with decreasing temperature along the giant branch was
found.  In ground-based work \citep{henry00}, a similar trend was found,
but the photometry was not precise enough ($\sim 1$ mmag) to see the
variability of the hotter G- and early-K giants ($\lesssim0.5$ mmag). The
timescales of the variations in these works were found to be inconsistent
with rotational modulation of a spotted photosphere, and were found to be
more consistent with acoustic oscillations of the atmospheres, with the
variations of the late-K and M giants consistent with radial pulsations,
and the variations of the more stable G and early-K giants dominated by
non-radial pulsations.

\begin{figure}

   \includegraphics[angle=0,scale=1.0,keepaspectratio=true]{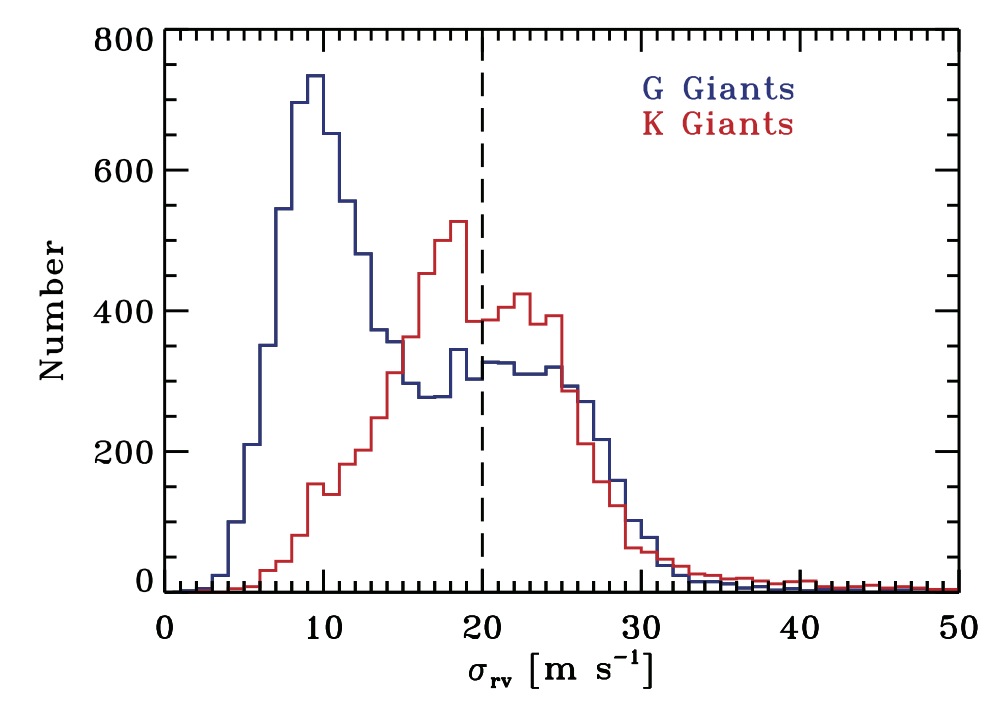}

    \figcaption{Distribution of radial velocity oscillations of G (blue)
    and K (red) giants predicted from the photometric dispersion and
    effective temperature \citep{kb95}.  The dashed line marks the median
    radial velocity oscillation for the K-giant sample of
    \citet{frink01}.\label{rvdist-fig}}

\end{figure}

Assuming that the photometric dispersion in the Kepler giants is also
dominated by acoustic oscillations, the photometric variations can be used
to predict radial velocity amplitudes of the oscillations. \citet{kb95}
developed a calibrated relationship between the velocity of oscillations
and the photometric amplitude variations:
\[
\sigma_{rv} = \left (
\frac{(\Delta F/F)_\lambda}{20.1\times 10^{-6}} \right ) \left (
\frac{\lambda}{0.55\micron}\right ) \left ( \frac{T_{eff}}{5777}\right )^2
\ \rm{m\ s^{-1}},
\]
where $\sigma_{rv}$ is the oscillation velocity of the star,  $(\Delta
F/F)_\lambda$ is the photometric flux change at the observed wavelength
$\lambda$, and $T_{eff}$ is the effective temperature of the star.  Using
this relation, we have calculated the expected radial velocity oscillations
for the G- and K-giants based upon their photometric dispersions and
effective temperatures (see Figure \ref{rvdist-fig}).

The bulk of predicted radial velocity dispersions are centered around
$10-20$ m/s with 90\% of the velocities $\lesssim 30$ m/s.   The K-giants
have a symmetric distribution centered at $\langle \sigma_{rv} \rangle
\approx 20\pm5$ m/s.  This is in good agreement with a radial velocity
study of K-giants \citep{frink01} where it was found that the radial
distribution of K-giants could be described with a Gaussian of mean 20 m/s
and width of 11 m/s with a long tail to higher velocity dispersions.  The
agreement with the predicted and measured distributions for representative
samples of K-giants suggests that the variability observed by Kepler is
dominated by acoustic oscillations in the atmospheres of the giants.

The G-giants predicted velocities show a bi-modal structure with peaks near
10 and 20 m/s  with the stronger peak towards lower radial velocity
variations.   The magnitude distributions of the G-giants that have
predicted radial velocity amplitudes of $<15$ m/s and those that have
predicated radial velocity amplitudes of $>15$ m/s are indistinguishable
indicating that the bimodality is not related to the brightness (and hence,
the photometric precision) of the stars, but rather is intrinsic to the
sample.  The radial velocity appears uncorrelated with temperature, but
does appear to have a weak anti-correlation with surface
gravity\footnote{The Kendall-$\tau$ non-parametric rank correlation value
between the surface gravities and the predicted radial velocity
oscillations is $-0.75$ (number of standard deviations from zero  is
$\approx 100$); a value of $-1$ would indicate a perfect
anti-correlation.}, suggesting  that the G-giant sample may contain a
sampling of dwarfs and sub-giants, which are atmospherically more stable
than the G-giants.

\subsubsection{Time Dependent Variability}\label{timevar-sec}

The analysis, thus far, has been performed on the full 33.5 day time
baseline of the quarter-1 dataset (30-minute cadence), but in reality,
stars are variable on a variety of timescales depending on the source of
the variability \citep[e.g.,  flares, pulsations, rotation, and
eclipses;][]{em08}.  A full detailed study of the variability as a function
of the time baseline and sampling rate is beyond the intent and scope of
this paper, but we have briefly explored how the variability fractions
change depending upon the length of the dataset investigated.  It should be
noted (as discussed above) that the Kepler public product may remove
long-term variability or enhance some forms of variability
\citep{vancleve10}, and the detailed results of this short study should be
viewed with that in mind.

\begin{figure}

	\includegraphics[angle=90,scale=0.7,keepaspectratio=true]{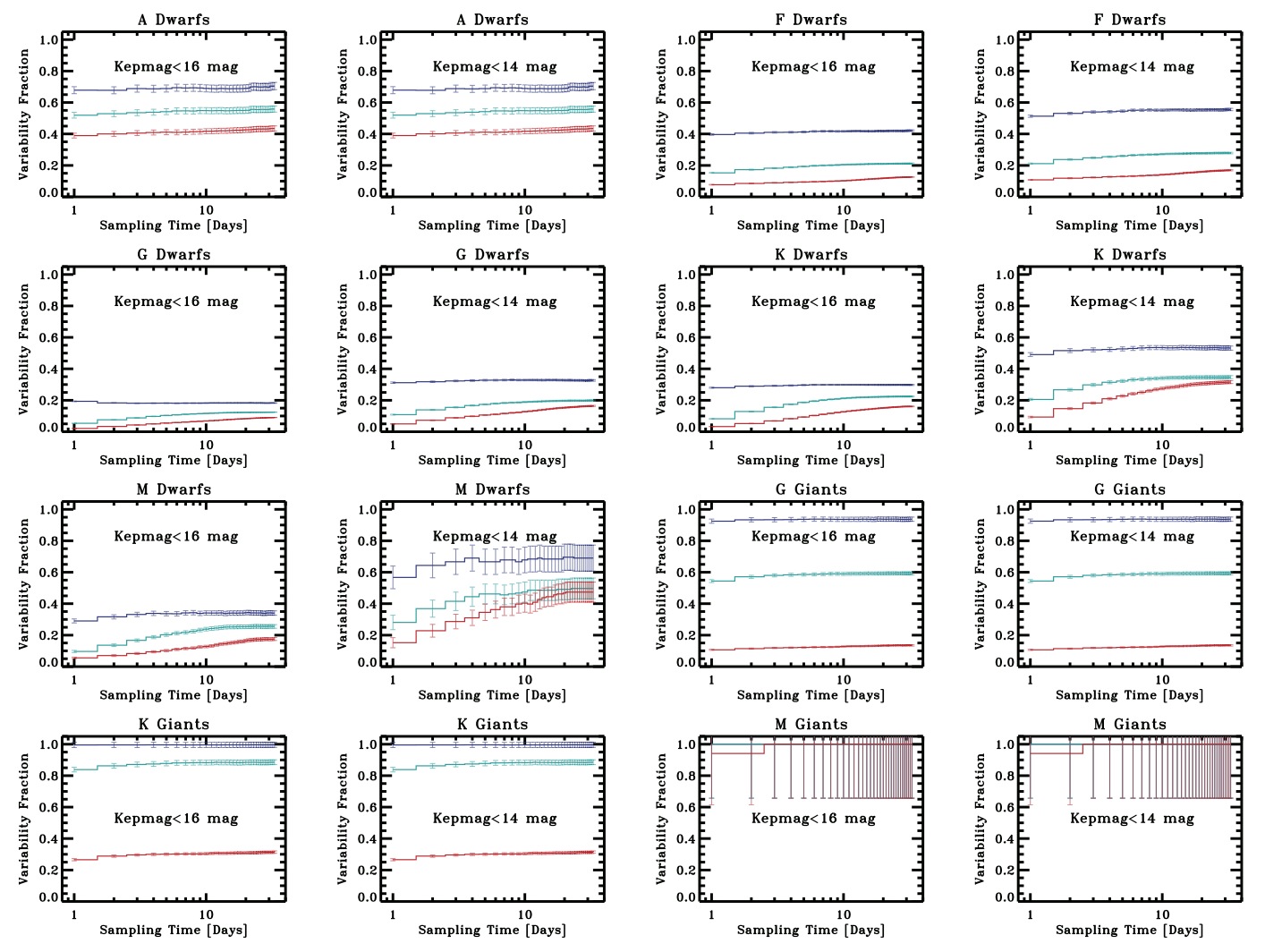}
    \vspace*{0.4in}

    \figcaption{Variability fraction distributions as a
    function of  sampling.   Each panel represents a different group of
    stars. There are two panels for each group; one panel for all the stars
    in the sample, and one panel where the stars were restricted to a
    Kepler magnitude of 14 or brighter. The contaminating giants in the
    M-dwarf sample and the contaminating dwarfs in the M-giant sample have
    been removed from the statistics.  The blue curves represent the
    fractions of stars with $\chi^2_\nu > 2$; the green curves represent
    the fractions of stars with $\chi^2_\nu > 10$;  the red curves
    represent the fractions of stars with $\chi^2_\nu >
    100$.\label{varfractime-fig}}

\end{figure}

For each of the light curves, we have assessed the light curve properties
with progressively longer time baselines starting at 1 day and extending to
33 days.   The median, the dispersion about the median, and the reduced
chi-square assuming a constant median value were calculated for each time
interval. To help alleviate biases that might arise from  sampling the
light curves in progressively longer samples in a single direction, the
statistics were calculated by sampling the light curves in the time-forward
direction ($0-1$ day, $0-2$ day $0-3$ day, $\ldots$ $0-33$ day) and in the
time-backward direction ($33-32$ day, $33-31$ day $33-30$ day, $\ldots$
$33-0$ day) and the results were averaged.  In Table \ref{var-tab} and
Figure \ref{varfractime-fig}, the dependency of the derived variability
fractions are summarized.  As with the overall variability statistics, the
analysis was performed for all stars (Kepmag $< 16$ mag) and for the
brighter end  of the sample (Kepmag $<14$ mag).

The overall fractions of giants that are variable ($\chi^2_\nu >2$) do not
change to within the  uncertainties of the fractions as the time baseline
is increased from 1 day to 33 days.  The fraction of giant stars that are
more significantly variable ($\chi^2_\nu >10$ and $\chi^2_\nu >100$) does
grow by $\approx 4-5\%$ from the 1-day baseline to the 33-day baseline.
The small growth of the variability fractions is likely a result of the
fact that nearly all of the giants are observed to be variable at the
precision of Kepler and the variability fraction has little room to change.

For the dwarf stars, the overall variability fractions ($\chi^2_\nu >2$)
increase by $\approx 1-5\%$, as the baseline is increased to 33 days.  As
with the giants, the variability fraction changes more substantially for
those stars that are more significantly variable ($\chi^2_\nu > 10$ and
$\chi^2_\nu > 100$).  Larger amplitude variability requiring longer time
periods is not surprising and has been observed previously
\citep[e.g.,][]{em08}. Increasing the time baseline from 1 day to 33 days
increases the variability fractions for the M-dwarf, K-dwarf, G-dwarf stars
variability more than for the F-dwarf and A-dwarf stars. This suggests that
the lower mass stars are predominately characterized by variability with
timescales of weeks (e.g., rotational modulation) while the higher mass
stars are predominately characterized by variability with timescales of
days (e.g., pulsations).

\subsection{Galactic Distribution}\label{glat-sec}

The Kepler field spans approximately 12 degrees in galactic latitude ($b
\approx 8^\circ-20^\circ$). Over this range of latitude, the different
galactic populations may play a role in the variability fractions. Because
the target samples are mostly magnitude-limited, the differing intrinsic
brightnesses of the stars lead to differing median distances of the stars
for each sub-group, and hence, to differing median heights ($z$) above the
galactic plane for a given line of sight.  \citet{walkowicz10} found a
higher fraction of the flaring M and K dwarfs at lower $z$-heights and they
suggested that they were sampling primarily the young thin disk.  Their
work inspired us to try to understand the overall variability fraction of
the sample as a function of latitude and $z$-height for each of the stellar
sub-groups.

\begin{figure}

    \includegraphics[angle=0,scale=0.5,keepaspectratio=true]{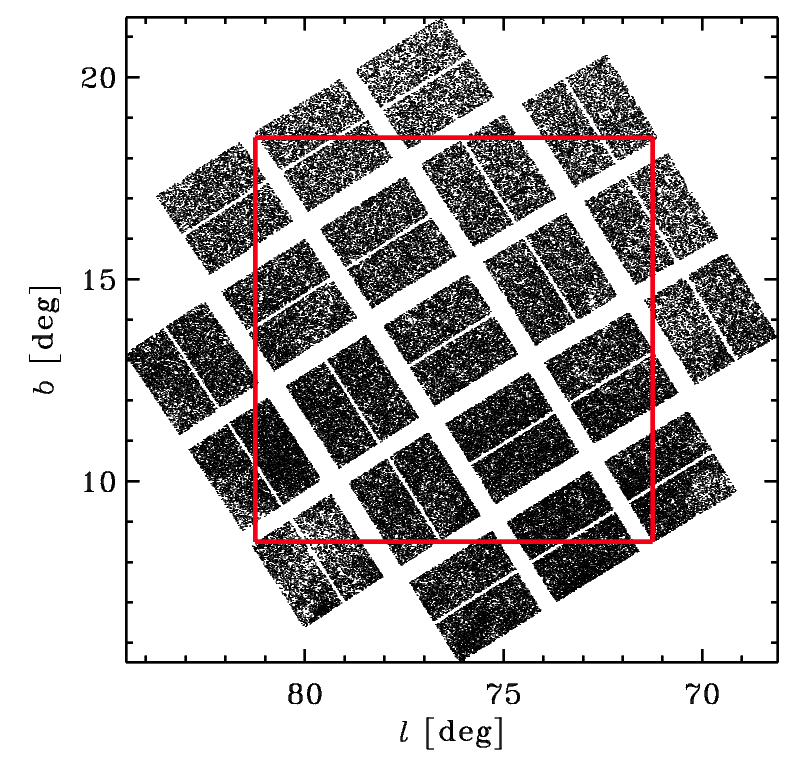}

    \figcaption{Galactic coordinates plot of the positions of all the stars
    in the sample.  The red-box delineates the $10^\circ \times 10^\circ$
    region used to explore the variability as a function of galactic
    latitude. \label{glatpos-fig}}

\end{figure}

\begin{figure}

    \includegraphics[angle=0,scale=0.8,keepaspectratio=true]{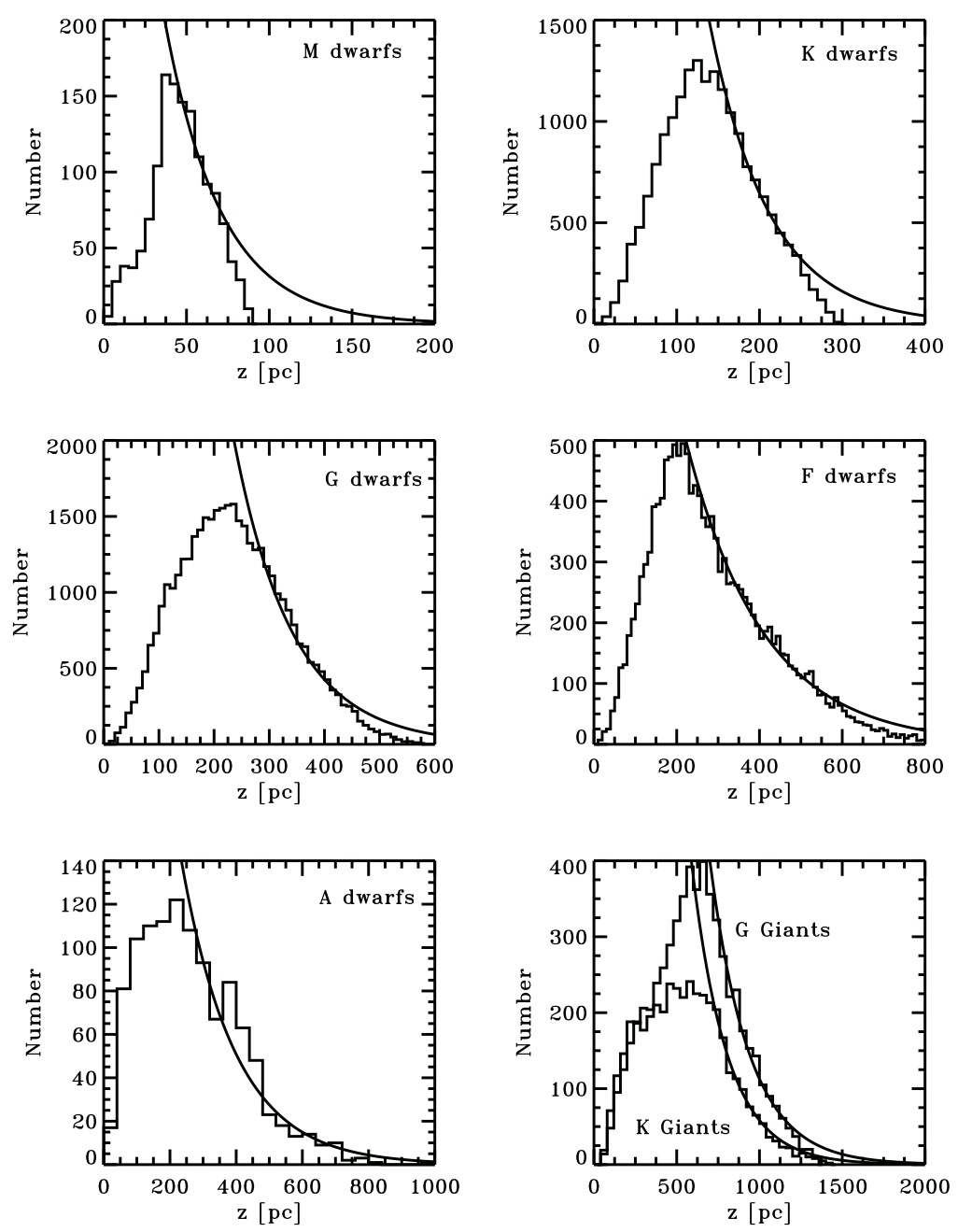}

    \figcaption{$z$-height distributions for the dwarfs and G and K
    giants.  The black smooth curves represent the best fit exponential
    curves to the distributions. The M-giants have been excluded from the
    plot because of the low number (23) in the sample. \label{zdist-fig}}

\end{figure}

A subset of the Kepler Field was selected (see Figure \ref{glatpos-fig}) to
remove the effects of the rotation of the Kepler field with respect to the
Galactic plane.  The median temperature and magnitude for each category of
stars was used to determine a ``typical'' distance for the stars, assuming
zero attenuation by interstellar dust (see Table \ref{height-tab}). The
$z$-height of each star was computed from the typical distance for its
sub-group, its apparent magnitude, and its galactic latitude.  This simple
estimation assumes that each star within a subgroup has the same absolute
magnitude.  While this, of course, is not strictly correct, the typical
spread of absolute magnitude within a sub-group is $\approx 1-2$ mag,
corresponding to only a factor of $1.2 -1.5$ in the distance.  The
$z$-height distributions of the stars (Figure~\ref{zdist-fig}) mostly
follow the expected exponential decay for a disk of the form $N \propto
\exp (-z/z_\circ)$ where $z_\circ$ is the characteristic scale height of
the disk \citep{crd96,juric08}.  Each $z$-height distribution was fitted
with a decaying exponential and the resulting scale heights are listed in
Table \ref{height-tab}.  The exponential fits work best for the
intrinsically brightest stars (e.g., the F-dwarfs, A-dwarfs, K-giants and
G-giants) where local distribution effects are minimized.

The M-dwarfs and K-dwarfs, with distances of only a few hundred parsecs and
scale heights of $z < 100$ pc, are dominated by stars located nearer to the
disk plane and by stars within the solar neighborhood.  The G-, F-, and
A-dwarfs all display larger characteristic scale heights ($z =  100 - 180$
pc), but are all within the expected size of the young thin disk. The K-
and G-giants have scale heights of $z = 200 - 250$pc which is
characteristic of the older thin disk \citep{juric08}.  If the stars came
from only this one disk population, it is expected that $\sim90\%$ of the
stars will have $z$-heights within $z \lesssim 2.3z_\circ$.  The actual
fractions are listed in Table~\ref{height-tab}; all of which are
significantly below 90\%, indicating that the thick disk may contribute to
the overall sample - particularly at higher galactic latitudes.  The thick
disk has a scale height of $\approx 900$ pc and a scaling fraction of $\sim
10\%$ \citep{juric08}.

\begin{figure}

    \includegraphics[angle=0,scale=0.8,keepaspectratio=true]{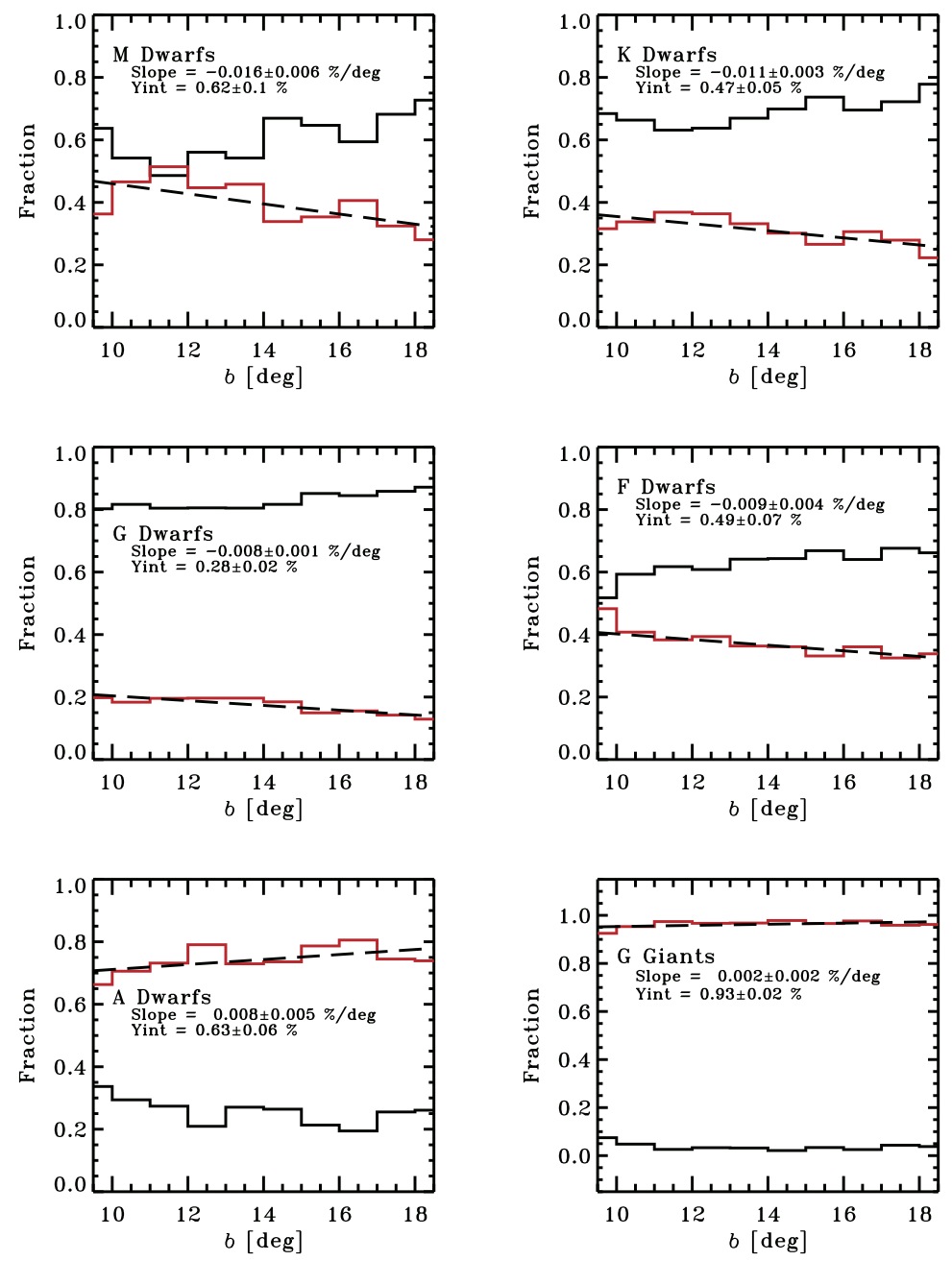}

    \figcaption{Galactic latitude distributions (binsize=1$^\circ$) for
    the dwarfs and G giants.  The black curves represent the fraction of
    stars within that galactic latitude bin that are deemed ``stable''
    ($\chi^2_\nu < 2$), and the red curves represent the fraction of those
    stars that are deemed ``variable'' ($\chi^2_\nu > 2$).  The black
    dashed line is a best fit to the variability fraction as a function of
    galactic latitude with the parameters of the line fit given in each
    panel. \label{glatvar-fig}}

\end{figure}

If the thin disk contributes only a portion (albeit the majority fraction)
to the sample of stars observed by Kepler, a variation in the variability
fraction as a function of galactic latitude (i.e., scale height) might be
expected.  Figure~\ref{glatvar-fig} displays the fraction of stable stars
($\chi^2_\nu < 2$) and variable stars ($\chi^2_\nu > 2$) as a function of
galactic latitude for each of the sub-groups (K-giants and M-giants are not
included in this sample as ``all'' of the K-giants are variable at the
precision of Kepler).  The M, K, G, and F dwarfs all show an increase in
the variability fraction as the galactic latitude gets lower (i.e., closer
to the plane). Moving higher in galactic latitude, the variability
fractions decrease by $\sim 0.1$\% over the $10^\circ$ span of the Kepler
Field.  This could indeed be the result of sampling younger stars in the
plane at lower latitudes as young stars are expected to be more active
\citep{west08}. Indeed, the flaring rate of M-dwarfs as a function of
$z$-height suggests that stars located nearer to the galactic plane are
more active and, hence, more variable \citep{walkowicz10}, in reasonable
agreement with what is discussed here.

An alternative explanation is that the background contamination is higher
when looking closer to along the galactic plane and that the increased
variability is the result of more significant blending of the primary star
with fainter background stars.  The slope of the variability fraction as a
function of latitude is strongest for the low luminosity stars (M- and
K-dwarfs), weakens as the intrinsic luminosity of the stars increases (G-
and F-dwarfs), and is not apparent for the most intrinsically bright stars
(A-dwarfs and G-giants).  As the Kepler sample is magnitude limited with
similar magnitude ranges for each of the stellar sub-groups, the different
sub-groups are essentially sampling different distances (see
Table~\ref{height-tab}).

For the thin disk ($z_\circ \sim 300$ pc), the path length to outside the
disk ($z \sim 600$ pc) is $\approx 4300$ pc at $b \sim 8^\circ$, but only
$\approx 1750$ pc at $b \sim 20^\circ$.  For the M-dwarfs with typical
distances of 200 pc, the latitude-change corresponds to a background  path
length (for a conic volume) difference of nearly 40\% from low ($b \sim
8^\circ$) to high ($b\sim20^\circ$) galactic latitude. For the G-, F-, and
A-dwarfs ($d \sim 1000$ pc), the background volume difference is $\lesssim
20\%$. The reduction in background path length is approximately 50\% from
M-dwarfs to F-dwarfs, which is also the fraction by which the slopes of the
variability fraction vs latitude change from M-dwarfs to G- and F-dwarfs
(see Figure \ref{glatvar-fig}).  The A-dwarfs do not display a reduction in
the variability fraction at higher latitudes; if anything, they exhibit a
weak (and somewhat insignificant) increase in variability at higher
latitudes. The G-giant stars, with typical distances that are larger than
the line of sight distances to the ``top'' of the exponential disk at $b
\sim 10-20^\circ$, show no dependence of the variability fraction on the
galactic latitude.  All of this is consistent with background stars
contributing to the variability of the primary stars.

Without a full model of the stellar galactic distribution {\it coupled}
with {\it a priori} knowledge of the true variability fraction of the
relative  populations, it is difficult to disentangle these scenarios
(true variability fractional changes as a function of latitude vs. changes
in the background contamination rate).  However, the apparent correlation
of flare rates with lower $z$-height \citep{walkowicz10} does suggest that
the higher variability fraction at lower galactic latitudes may be real and
the result of sampling a systematic younger population.

\subsection{M-Dwarf Variability}\label{mdwarf-sec}

M-dwarfs are favorable targets to search for earth-sized planets because
the transits are relatively deep ($\sim 1-3$ mmag), and the radial velocity
signatures are relatively large ($\sim 10$ m/s).  In addition, planets in
the habitable zones of M-stars are in relatively short orbits ($10-20$
days) compared to that of the habitable zones for sun-like stars ($\sim 1$
year).  As a result there has been a strong interest in the community for
searching for planets around M-stars \citep{irwin09,charbonneau09,bean10}.
Thus, understanding the M-dwarf variability amplitudes and fractions is
critical to understanding how complete such transit and radial velocity
surveys can be.

\begin{figure}

   \includegraphics[angle=0,scale=0.5,keepaspectratio=true]{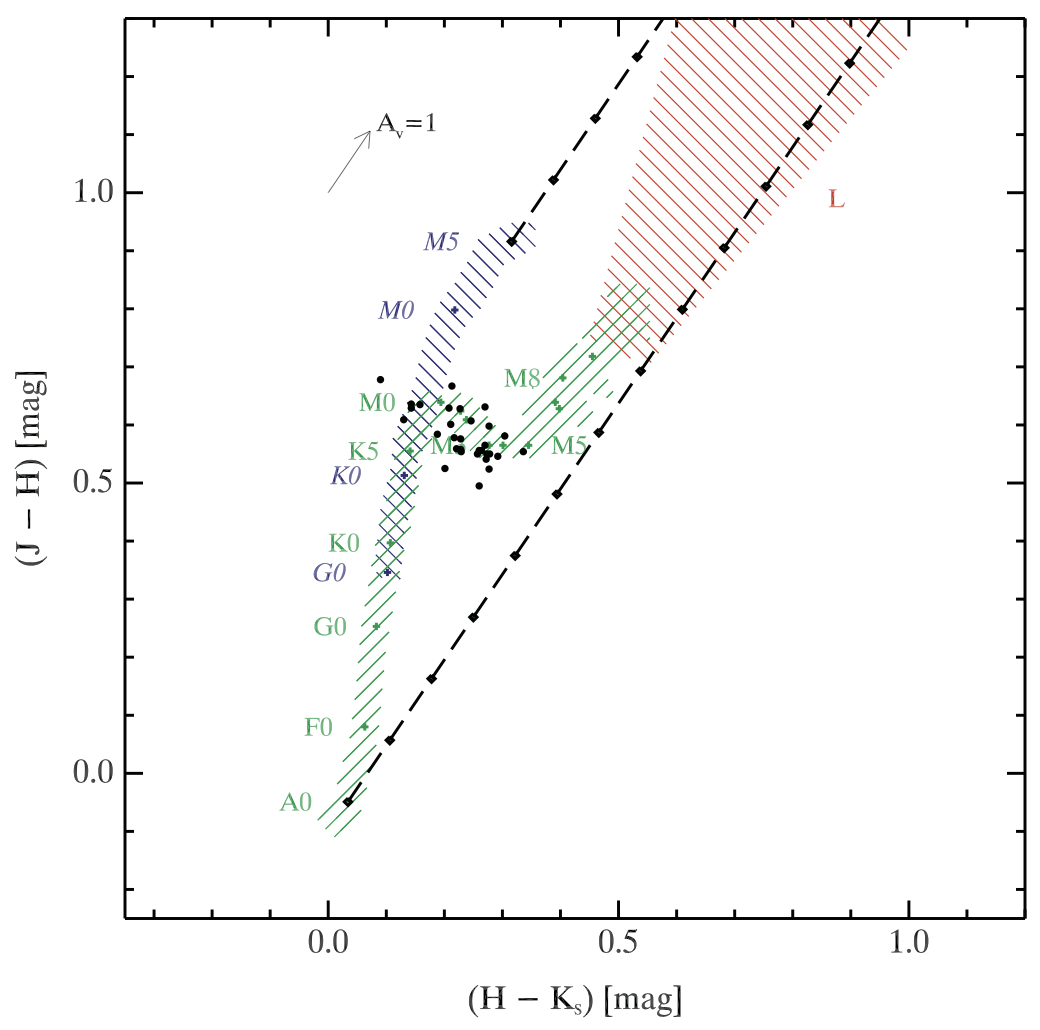}

    \figcaption{2MASS color-color diagram for the 29 stars identified as
    M-dwarfs from outside catalogs. \label{knownjhk-fig}}

\end{figure}

In the previous sections (\S\ref{dispersion-sec},\ref{fraction-sec}), the
overall variability fraction of the M-dwarfs was found to be $\sim 40-70\%$
with dispersions of $\sigma_m \sim 3-5$ mmag, depending on the brightness
of the stars being considered.    As an alternative, in this section we
identify a small sample of relatively bright, certain M-dwarfs based on
well-vetted proper motion catalogs and analyze their variability in more
detail.  These M dwarfs include all of the M dwarfs in the Kepler field
(with Q1 light curves) from the Gliese and LHS catalogs
\citep{stauffer10}, and the brightest stars in the LSPM catalog with $V-J
> 2.6$ (i.e. colors consistent with an M dwarf).  A plot of $J-H$ vs $H-K$
confirms that these are indeed M dwarfs (Figure~\ref{knownjhk-fig}).  Only
four of these stars have KIC $T_{eff}$ or  $\log(g)$ -- the rest would be
absent from statistical studies which rely on $T_{eff}$ and $\log(g)$ to
identify dwarfs, and, of these, one would have been classified as a giant.

\begin{figure}

    \includegraphics[angle=90,scale=0.5,keepaspectratio=true]{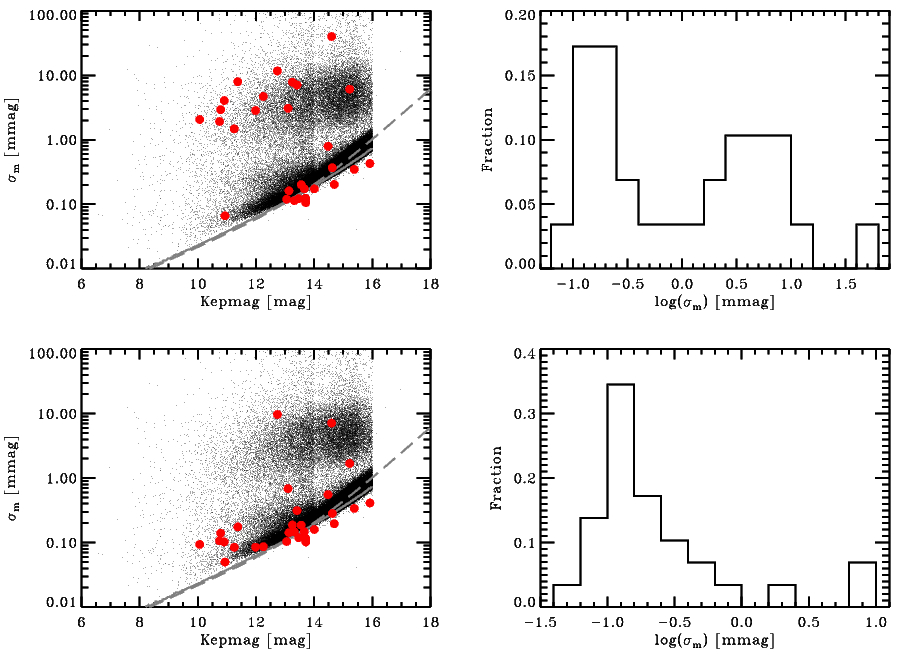}

    \figcaption{Left panels show the photometric dispersion plotted as a
    function of magnitude for the KIC-identified M-dwarfs with a color
    restriction of J-H$<0.75$ mag (black dots) and for the outside
    identified M-dwarfs (red).  The solid grey line represents the median
    uncertainty as reported in the Kepler data product.  The dashed grey
    curve is the uncertainty upper limit curve from \citet{jenkins10}.
    Right panels show the distributions of the (logarithmic) photometric
    dispersion (binsize = 0.2 dex) for the known M-dwarfs (red points in
    left figures). The top panels reflect the dispersion of the known
    M-dwarfs determined for the entire light curve (30 days); the bottom
    panels reflect the dispersion calculated from the point-to-point
    differences on 12-hour timescales (only for the known (red points)
    M-dwarfs). The KIC-identified M-dwarfs (black dots) are shown at the
    30-minute cadence dispersion in both plots for
    reference.\label{knowndisp-fig}}

\end{figure}

To understand the variability of these bright M-dwarfs on the time scales
relevant to planetary transits, we have calculated the short-term 12-hour
variability for each of the light curves, by computing the dispersions in
running 12-hour time bins.  The median of all the 12- hour bin dispersions
for each light curve were calculated and taken as representative of the
12-hour variability timescales for the M-dwarfs.  The dispersions for the
full time series (33 day) and for the 12-hour timescales are listed in
Table~\ref{mdwarf-tab}. In all cases, the dispersion on the 12-hour
timescale is smaller than the full 30 day dispersion, and for many of the
stars, the dispersion drops to the photometric limit of instrument (see
Figure~\ref{knowndisp-fig}).

On the 33-day timescale, the dispersion is bimodal with peaks near 0.1 mmag
and 5 mmag.  The  0.1 mmag peak is dominated by stars which are quiet to
the precision of the instrument, and $\approx 1/2$ (15/29) of the sample
are variable with $\chi^2_\nu > 2$ and dispersions of $\sigma_m \gtrsim 1$
mmag.  For the 12-hour timescale, the variability fraction drops
significantly with only 6 stars that have dispersions $\sigma_m > 0.5$
mmag. The high dispersion is likely caused by rotational variability with
periods of 1 day or longer; thus, it is not all that surprising that the
dispersion drops when the light curves are sampled at 12-hour timescales.
A prime example of this is LHS6343 (KIC 10002261) which is a newly
discovered transiting brown dwarf \citep{johnson10}.  The dispersion for
the entire light curve is $\approx 3$ mmag, but the 12-hour timescale
dispersion matches the out-of-eclipse dispersion of $\approx 0.7$ mmag -
much like what is observed for the transiting planets around FGK stars (see
\S~\ref{dispersion-sec}).

\section{Summary}\label{summary-sec}

An analysis of the variability statistics of the stars in the Quarter-1
publicly released Kepler data has been performed.  The Kepler data cover
33.5 days and are sampled at a 30 minute cadence.  The Kepler Input Catalog
parameters have been used to separate the 150,000 stars into dwarfs and
giants which were further separated into temperature bins corresponding
roughly to spectral classes A, F, G, K and M.

The majority of the dwarf stars were found to be photometrically quiet down
to the per-observation (30 minute) precision of the Kepler spacecraft.  The
derived variability fractions range from 10 - 100\% depending on the
stellar group and brightness range explored. The G-dwarfs are the most
stable with $<20\%$ of the all the stars in the sample having a $\chi^2_\nu
\gtrsim 2$.  The G-dwarfs appear to have a dispersion noise floor of $\sim
0.04$ mmag for the 30-minute sampling of the Kepler data.

At the precision of Kepler, $>95$\% of K, G, and M giants are variable with
noise floors of $\sim 0.1$ mmag, $\sim 0.3$ mmag, and $\sim 10$ mmag,
respectively. The photometric dispersion of the giants is consistent with
acoustic variations of the photosphere. The photometrically-predicted
radial velocity distribution for the K-giants is in agreement with the
measured distribution; the G-giant radial velocity distribution is bimodal
which may indicate a transition from sub-giant to giant.

We also briefly explored the dependence of the variability fractions as a
function of time baseline of the light curves.  In general, increasing the
length of the light curve baseline increased the fraction of stars that are
variable.  For the dwarf stars, the lower mass stars were found to be
predominately characterized by variability with timescales of weeks (e.g.,
rotational modulation) while the higher mass stars were found to be
predominately characterized by variability with timescales of days (e.g.,
pulsations).  For the giant stars, the variability fractions changed very
little from a 1-day sampling to a 33-day sampling.

A study of the distribution of the variability as a function of galactic
latitude suggests sources closer to the galactic plane are more variable.
The scale height distribution of the dwarfs is consistent with the young
thin disk, and the scale height of the giants is consistent with the older
thin disk.  For the lower mass stars (M, K, and G dwarfs), the variability
fraction decreases with increasing galactic latitude. This may be the
result of sampling differing populations as a function of latitude and
preferentially sampling younger stars at lower galactic latitudes within
the Kepler field.

In addition to the statistical study of M dwarf variability using the ~2500
relatively anonymous probable M dwarfs in the Kepler field, we have also
examined the variability of 29 known M dwarfs in the Kepler field drawn
from the GJ, LHS, and LSPM catalogs. The analysis of the known M dwarfs
indicates that the M dwarfs are primarily variable on timescales of weeks
presumably dominated by spots, rotation, and binarity.  But on shorter
timescales of hours-to-days, the stars are quieter by nearly an order of
magnitude. At these shorter timescales, the variability fraction of the
M-dwarfs drops from $\sim40\%$ to $\sim20\%$.  The shorter timescales are
relevant for searches of planetary transits which typically last a few
hours.  In general, a search for transiting earth-sized planets around
M-stars should not be hampered by the typical stellar variability of
M-dwarfs.

\acknowledgments  The authors would like to acknowledge the referee for his
or her extremely insightful and useful comments which made this a better
paper. Portions of this work were performed at the California Institute of
Technology under contract with the National Aeronautics and Space
Administration. This research has made use of the NASA/IPAC Star and
Exoplanet Database, which is operated by the Jet Propulsion Laboratory,
California Institute of Technology, under contract with the National
Aeronautics and Space Administration.

\newpage

\begin{deluxetable}{ccrcr}
\tablecolumns{5}
\tablewidth{5in}
\tablecaption{KIC-Based Temperature Bins\label{temp-tab}}
\tablehead{
\colhead{Spectral} & \colhead{Dwarf} & \colhead{Dwarf} &
\colhead{Giant}&  \colhead{Giant}\\
\colhead{Type} & \colhead{T$_{eff}$Range} & \colhead{Number\tablenotemark{a}} &
\colhead{T$_{eff}$Range} & \colhead{Number\tablenotemark{a}}
}
\startdata
A & $>7300$         &  2311  (2296) & \nodata        & 0    \\
F & $(6000 - 7300]$ & 23750 (15996) & \nodata        & 0    \\
G & $(5300 - 6000]$ & 66682 (17940) & $> 4800 $      & 9880 (9877)\\
K & $(4000 - 5300]$ & 30889  (4874) & $(3800, 4800]$ & 7226 (7225)\\
M & $\leq 4000$     &  2460   (171) & $\leq 3800$    &   23   (17)
\enddata
\tablenotetext{a}{Number is parentheses is number of stars brighter
than Kepmag$ < 14$ mag.}
\end{deluxetable}

\begin{deluxetable}{rrrrrrr}
\tablecolumns{7}
\tablewidth{6.75in}
\tablecaption{Stars Separated by Class, Magnitude, and Dispersion\tablenotemark{a}
\label{disp-tab}}
\tablehead{
\colhead{Stellar} & \colhead{Kepler} & \colhead{\# of Stars} & \colhead{\# of Stars} &
\colhead{\# of Stars} & \colhead{\# of Stars} & \colhead{\# of Stars}\\
\colhead{Group} & \colhead{Magnitude} & \colhead{in Magnitude} & \colhead{with} &
\colhead{with} &
\colhead{with} & \colhead{with}\\
& \colhead{Range} & \colhead{Range} & \colhead{$\sigma < 0.1$} &
\colhead{$\sigma = 0.1 - 1$} &
\colhead{$\sigma = 1 - 10$} & \colhead{$\sigma > 10$}
}
\startdata
M Dwarfs\tablenotemark{b}  & $<$10  &     3  &     0  &       0 &    3 &   0 \\
         & 10-12  &    13  &     2  &       3 &    8 &   0 \\
         & 12-14  &   154  &     0  &      83 &   57 &  14 \\
         & 14-16  &  2182  &     0  &    1503 &  529 & 150 \\
K Dwarfs & $<$10  &    18  &     3  &      10 &    3 &   2 \\
         & 10-12  &   264  &    75  &      83 &   94 &  11 \\
         & 12-14  &  4588  &    92  &    3063 & 1212 & 221 \\
         & 14-16  & 26019  &     1  &   19892 & 5184 & 942 \\
G Dwarfs & $<$10  &    63  &    26  &      27 &   10 &   0 \\
         & 10-12  &  1297  &   716  &     257 &  294 &  27 \\
         & 12-14  & 16566  &   542  &   13376 & 2332 & 316 \\
         & 14-16  & 48756  &     0  &   43533 & 4550 & 673 \\
F Dwarfs & $<$10  &   141  &    28  &      81 &   29 &   3 \\
         & 10-12  &  1948  &   490  &     966 &  429 &  54 \\
         & 12-14  & 13906  &   402  &   11407 & 1806 & 291 \\
         & 14-16  &  7755  &     0  &    7158 &  519 &  78 \\
A Dwarfs & $<$10  &   231  &   114  &      56 &   56 &   5 \\
         & 10-12  &   739  &   280  &     210 &  226 &  20 \\
         & 12-14  &  1326  &   124  &     658 &  460 &  84 \\
         & 14-16  &    15  &     0  &       5 &    8 &   2 \\
M Giants\tablenotemark{c}  & $<$10  &     6  &     0  &       0 &    3 &   3 \\
         & 10-12  &     8  &     0  &       0 &    4 &   4 \\
         & 12-14  &     3  &     0  &       0 &    1 &   2 \\
         & 14-16  &     0  &     0  &       0 &    0 &   0 \\
K Giants & $<$10  &   350  &     0  &     273 &   73 &   4 \\
         & 10-12  &  1683  &     1  &    1468 &  200 &  11 \\
         & 12-14  &  5192  &     1  &    4776 &  349 &  66 \\
         & 14-16  &     0  &     0  &       0 &    0 &   0 \\
G Giants & $<$10  &   233  &     5  &     209 &   17 &   2 \\
         & 10-12  &  1619  &    38  &    1467 &   93 &  21 \\
         & 12-14  &  8025  &     7  &    7689 &  255 &  74 \\
         & 14-16  &     0  &     0  &       0 &    0 &   0
\enddata
\tablenotetext{a}{Dispersions in milli-magnitudes [mmag].}
\tablenotetext{b}{The 108 contaminating giants in the M-dwarf sample have been
    removed from the statistics.}
\tablenotetext{c}{The 6 contaminating dwarfs in the M-giant sample have been
    removed from the statistics.}
\end{deluxetable}

\begin{deluxetable}{rlcccccc}
\tablecolumns{8}
\tablewidth{7in}
\tablecaption{Variability Fractions\tablenotemark{a} \label{var-tab}}
\tablehead{
\colhead{Time} &  &
\multicolumn{2}{c}{$\chi^2_\nu > 2$} &
\multicolumn{2}{c}{$\chi^2_\nu > 10$} &
\multicolumn{2}{c}{$\chi^2_\nu > 100$}\\
\colhead{scale}& \colhead{Category} & \colhead{$\le 16$ mag} &
\colhead{$\le 14$ mag} &
\colhead{$\le 16$ mag} & \colhead{$\le 14$ mag} &
\colhead{$\le 16$ mag} & \colhead{$\le 14$ mag}
}
\startdata
33 day & All Dwarfs & 0.269 (0.002) & 0.461 (0.004) & 0.177 (0.001) & 0.269 (0.003) & 0.123 (0.001) & 0.200 (0.002)\\
& M Dwarfs\tablenotemark{b}  & 0.367 (0.014) & 0.690 (0.083) & 0.285 (0.012) & 0.497 (0.065) & 0.209 (0.010) & 0.474 (0.064)\\
& K Dwarfs   & 0.298 (0.004) & 0.532 (0.013) & 0.224 (0.003) & 0.347 (0.010) & 0.161 (0.002) & 0.316 (0.009)\\
& G Dwarfs   & 0.183 (0.002) & 0.325 (0.005) & 0.125 (0.002) & 0.199 (0.004) & 0.090 (0.001) & 0.164 (0.003)\\
& F Dwarfs   & 0.421 (0.005) & 0.555 (0.007) & 0.212 (0.003) & 0.279 (0.005) & 0.127 (0.003) & 0.169 (0.004)\\
& A Dwarfs   & 0.705 (0.023) & 0.705 (0.023) & 0.559 (0.019) & 0.558 (0.019) & 0.434 (0.016) & 0.435 (0.016)\\
& All Giants & 0.962 (0.011) & 0.962 (0.011) & 0.717 (0.009) & 0.717 (0.008) & 0.210 (0.004) & 0.210 (0.004)\\
& M Giants\tablenotemark{c}  & 1.000 (0.343) & 1.000 (0.343) & 1.000 (0.343) & 1.000 (0.343) & 1.000 (0.343) & 1.000 (0.343) \\
& K Giants   & 0.996 (0.017) & 0.996 (0.017) & 0.886 (0.015) & 0.886 (0.015) & 0.314 (0.008) & 0.314 (0.008)\\
& G Giants   & 0.938 (0.014) & 0.938 (0.014) & 0.593 (0.010) & 0.593 (0.010) & 0.135 (0.004) & 0.135 (0.004)\\
\\
10 day & All Dwarfs & 0.268 (0.002) & 0.461 (0.004) & 0.167 (0.001) & 0.260 (0.003) & 0.097 (0.001) & 0.167 (0.002)\\
& M Dwarfs\tablenotemark{b}  & 0.367 (0.014) & 0.678 (0.082) & 0.269 (0.012) & 0.474 (0.063) & 0.163 (0.009) & 0.404 (0.058)\\
& K Dwarfs   & 0.299 (0.004) & 0.535 (0.013) & 0.212 (0.003) & 0.341 (0.010) & 0.127 (0.002) & 0.275 (0.008)\\
& G Dwarfs   & 0.182 (0.002) & 0.328 (0.005) & 0.117 (0.001) & 0.189 (0.004) & 0.069 (0.001) & 0.128 (0.003)\\
& F Dwarfs   & 0.417 (0.005) & 0.551 (0.007) & 0.205 (0.003) & 0.272 (0.005) & 0.102 (0.002) & 0.141 (0.003)\\
& A Dwarfs   & 0.690 (0.022) & 0.689 (0.023) & 0.548 (0.019) & 0.548 (0.019) & 0.417 (0.016) & 0.417 (0.016)\\
& All Giants & 0.961 (0.011) & 0.962 (0.011) & 0.713 (0.009) & 0.713 (0.008) & 0.202 (0.004) & 0.202 (0.004)\\
& M Giants\tablenotemark{c}  & 1.000 (0.343) & 1.000 (0.343) & 1.000 (0.343) & 1.000 (0.343) & 1.000 (0.343) & 1.000 (0.343)\\
& K Giants   & 0.995 (0.017) & 0.995 (0.017) & 0.883 (0.015) & 0.838 (0.015) & 0.303 (0.007) & 0.304 (0.007)\\
& G Giants   & 0.937 (0.014) & 0.937 (0.014) & 0.589 (0.010) & 0.589 (0.010) & 0.128 (0.004) & 0.128 (0.004)\\
\\
1 day & All Dwarfs & 0.264 (0.002) & 0.432 (0.004) & 0.089 (0.001) & 0.183 (0.002) & 0.042 (0.001) & 0.097 (0.002)\\
& M Dwarfs\tablenotemark{b}  & 0.320 (0.013) & 0.567 (0.072) & 0.128 (0.008) & 0.281 (0.046) & 0.077 (0.006) & 0.152 (0.032)\\
& K Dwarfs   & 0.279 (0.003) & 0.491 (0.012) & 0.082 (0.002) & 0.205 (0.007) & 0.033 (0.001) & 0.093 (0.005)\\
& G Dwarfs   & 0.193 (0.002) & 0.312 (0.005) & 0.054 (0.001) & 0.105 (0.003) & 0.021 (0.001) & 0.050 (0.002)\\
& F Dwarfs   & 0.397 (0.005) & 0.513 (0.007) & 0.152 (0.003) & 0.211 (0.004) & 0.077 (0.002) & 0.108 (0.003)\\
& A Dwarfs   & 0.678 (0.022) & 0.679 (0.022) & 0.519 (0.018) & 0.520 (0.019) & 0.389 (0.015) & 0.340 (0.015)\\
& All Giants & 0.955 (0.010) & 0.955 (0.010) & 0.668 (0.008) & 0.668 (0.008) & 0.174 (0.003) & 0.174 (0.003)\\
& M Giants\tablenotemark{c}  & 1.000 (0.343) & 1.000 (0.343) & 1.000 (0.343) & 1.000 (0.343) & 0.941 (0.323) & 0.941 (0.328)\\
& K Giants   & 0.995 (0.017) & 0.995 (0.017) & 0.839 (0.015) & 0.838 (0.015) & 0.266 (0.007) & 0.266 (0.007)\\
& G Giants   & 0.925 (0.013) & 0.925 (0.013) & 0.544 (0.009) & 0.544 (0.009) & 0.107 (0.003) & 0.107 (0.003)

\enddata
\tablenotetext{a}{Values in parentheses are uncertainties
based upon the propagation of errors of the counting statistics.}
\tablenotetext{b}{The 108 contaminating giants in the M-dwarf sample have been
    removed from the statistics.}
\tablenotetext{c}{The 6 contaminating dwarfs in the M-giant sample have been
    removed from the statistics.}
\end{deluxetable}

\begin{deluxetable}{rccccc}
\tablecolumns{6}
\tablewidth{5in}
\tablecaption{Galactic Distributions \label{height-tab}}
\tablehead{
\colhead{} & \colhead{Median}  & \colhead{Median} & \colhead{Typical} &
\colhead{Scale} & \colhead{Fraction of}\\
\colhead{} & \colhead{Teff} & \colhead{Kepmag}  &
\colhead{Distance} & \colhead{Height $z_\circ$} & \colhead{Stars}\\
\colhead{Category} & \colhead{[K]} & \colhead{[mag]}  &
\colhead{[pc]} & \colhead{[pc]\tablenotemark{a}}& \colhead{$z \leq 2.3z_\circ$}
}

\startdata
M Dwarfs   &   3800 & 15.3 & 200  & 35  & 0.97\\
K Dwarfs   &   5000 & 15.1 & 600  & 75  & 0.68\\
G Dwarfs   &   5700 & 14.7 & 1000 & 105 & 0.56\\
F Dwarfs   &   6200 & 13.6 & 1100 & 185 & 0.83\\
A Dwarfs   &   8000 & 12.3 & 1200 & 165 & 0.77\\
G Giants   &   5000 & 13.1 & 2800 & 235 & 0.40\\
K Giants   &   4700 & 12.7 & 2500 & 215 & 0.43
\enddata
\tablenotetext{a}{Calculated by fitting the $z$-height distributions
in Figure~\ref{zdist-fig}.}
\end{deluxetable}

\begin{deluxetable}{rrcccc}
\tablecolumns{6}
\tablewidth{5in}
\tablecaption{M-Dwarf Stars \label{mdwarf-tab}}
\tablehead{
\colhead{Star}  & \colhead{KIC} & \colhead{KIC} & \colhead{KIC} &
\colhead{$\sigma_m$} & \colhead{$\sigma_m$} \\
\colhead{Name} & \colhead{ID} & \colhead{T$_{eff}$} & \colhead{$\log(g)$} &
\colhead{33 day}  & \colhead{12 hour}\\
\colhead{} & \colhead{} & \colhead{[K]} & \colhead{[cm s$^{-2}$]} &
\colhead{[mmag]}  & \colhead{[mmag]}
}

\startdata
LHS6351	      &	2164791  &	\nodata	&	\nodata	&	6.09	&	1.69\\
LSP1912+3826  &	3330684  &	\nodata	&	\nodata	&	0.42	&	0.41\\
LSP1909+3910  &	4043389  &	3713	&	4.385	&	7.94	&	0.17\\
GJ4099	      &	4142913  &	\nodata	&	\nodata	&	4.05	&	0.10\\
GJ4113	      &	4470937  &	\nodata	&	\nodata	&	0.06	&	0.04\\
LSP1917+4007  &	5002836  &	\nodata	&	\nodata	&	0.10    &	0.10\\
LSP1947+4020  &	5206997  &	\nodata	&	\nodata	&	2.93	&	0.13\\
LSP1935+4119  &	6049470  &	\nodata	&	\nodata	&	2.05	&	0.09\\
LSP1919+4127  &	6117602  &	\nodata	&	\nodata	&	3.84	&	2.13\\
LSP1858+4147  &	6345835  &	\nodata	&	\nodata	&	2.84	&	0.08\\
LSP1956+4149  &	6471285  &	3201	&	0.07	&	0.20    &	0.18\\
LSP1927+4231  &	7033670  &	\nodata	&	\nodata	&	0.36	&	0.28\\
LSP1944+4232  &	7049465  &	4033	&	4.505	&	1.47	&	0.08\\
LSP1912+4239  &	7106807  &	\nodata	&	\nodata	&	0.12	&	0.11\\
LSP1912+4316  &	7596910  &	\nodata	&	\nodata	&	0.20    &	0.19\\
LHS6349	      &	7820535  &	\nodata	&	\nodata	&	0.34	&	0.33\\
LSP1854+4447  &	8607728  &	\nodata	&	\nodata	&	7.06	&	0.31\\
LSP2001+4500  &	8846163  &	\nodata	&	\nodata	&	0.17	&	0.15\\
LHS3429	      &	8872565  &	\nodata	&	\nodata	&	0.12	&	0.11\\
LSP1933+4515  &	8957023  &	3553	&	4.117	&	7.82	&	0.18\\
LHS3420	      &	9201463  &	\nodata	&	\nodata	&	39.8	&	7.13\\
GJ1243        &	9726699  &	\nodata	&	\nodata	&	11.6	&	9.69\\
LHS6343	      &	10002261 &	\nodata	&	\nodata	&	3.07	&	0.68\\
LSP1857+4720  &	10258179 &	\nodata	&	\nodata	&	0.11	&	0.10\\
LSP1854+4736  &	10453314 &	\nodata	&	\nodata	&	0.17	&	0.14\\
GJ4083	      &	10647081 &	\nodata	&	\nodata	&	4.69	&	0.08\\
LSP1916+4949 &	11707868 &	\nodata	&	\nodata	&	0.11	&	0.14\\
LSP1948+5015 &	11925804 &	\nodata	&	\nodata	&	0.79	&	0.55\\
LSP1919+5130 &	12555642 &	\nodata	&	\nodata	&	1.92	&	0.10
\enddata
\end{deluxetable}

\clearpage
\clearpage

\clearpage

\clearpage

\clearpage

\clearpage

\clearpage

\clearpage

\clearpage

\clearpage

\clearpage

\clearpage

\clearpage

\clearpage

\clearpage

\clearpage

\end{document}